\documentclass{emulateapj}

\newcommand{\citepeg}[1]{\citep[{e.g.,}][]{#1}}
\def\Swift{{\textit{Swift}}\,}

\shorttitle{Host Galaxies of Swift Dark GRBs}
\shortauthors{Perley et al.}

\begin{document}

\title{The Host Galaxies of Swift Dark Gamma-Ray Bursts: Observational Constraints on Highly Obscured and Very High-Redshift GRBs}

\def\ucb{1}
\def\chicago{2}
\def\stanford{3}
\def\ucsc{4}
\def\cfa{5}
\def\keckpd{6}
\def\swinburne{7}
\def\caltech{8}
\def\chile{9}
\def\iofa{10}
\def\hubble{11}
\def\lcogt{12}
\def\mail{13}

\author{D.~A.~Perley\altaffilmark{\ucb,\mail},
        S.~B.~Cenko\altaffilmark{\ucb},
        J.~S.~Bloom\altaffilmark{\ucb},
        H.-W.~Chen\altaffilmark{\chicago},
        N.~R.~Butler\altaffilmark{\ucb},
        D.~Kocevski\altaffilmark{\stanford},
        J.~X.~Prochaska\altaffilmark{\ucsc},
        M.~Brodwin\altaffilmark{\cfa,\keckpd},
        K.~Glazebrook\altaffilmark{\swinburne},
        M.~M.~Kasliwal\altaffilmark{\caltech},
        S.~R.~Kulkarni\altaffilmark{\caltech},
        S.~Lopez\altaffilmark{\chile},
        E.~O.~Ofek\altaffilmark{\caltech},
        M.~Pettini\altaffilmark{\iofa},
        A.~M.~Soderberg\altaffilmark{\cfa,\hubble},
        D.~Starr\altaffilmark{\ucb,\lcogt}
        }

\altaffiltext{\ucb}{Department of Astronomy, University of California, Berkeley, CA 94720-3411.}
\altaffiltext{\chicago}{Department of Astronomy and Astrophysics, University of Chicago, 5640 S. Ellis Ave, Chicago, IL 60637}
\altaffiltext{\stanford}{Kavli Institute for Particle Astrophysics and Cosmology, Stanford University, 2575 Sand Hill Road M/S 29, Menlo Park, CA 94025}
\altaffiltext{\ucsc}{Department of Astronomy and Astrophysics, UCO/Lick Observatory; University of California, 1156 High Street, Santa Cruz, CA 95064}
\altaffiltext{\cfa}{Harvard-Smithsonian Center for Astrophysics, 60 Garden Street, Cambridge, MA 02138}
\altaffiltext{\keckpd}{W. M. Keck Postdoctoral Fellow at the Harvard-Smithsonian Center for Astrophysics}
\altaffiltext{\swinburne}{Swinburne University of Technology, PO Box 218, Hawthorn, Victoria 3122,
Australia}
\altaffiltext{\caltech}{Department of Astronomy, California Institute of Technology, M/C 249-17, Pasadena, CA 91125}
\altaffiltext{\chile}{Departamento de Astronom\'ia, Universidad de Chile, Casilla 36-D, Santiago, Chile.}
\altaffiltext{\iofa}{Institute of Astronomy, Madingley Road, Cambridge, CB3 0HA, UK}
\altaffiltext{\hubble}{Hubble Fellow}
\altaffiltext{\lcogt}{Las Cumbres Observatory Global Telescope Network, Inc.  6740 Cortona Dr. Suite 102, Santa Barbara, CA 93117}
\altaffiltext{\mail}{e-mail: dperley@astro.berkeley.edu}

\slugcomment{Submitted to AJ 2009-04-30; Accepted 2009-09-19; Published 2009-12}

\begin{abstract}
In this work we present the first results of our imaging campaign at Keck Observatory to identify the host galaxies of ``dark'' gamma-ray bursts (GRBs), events with no detected optical afterglow or with detected optical flux significantly fainter than expected from the observed X-ray afterglow.  We find that out of a uniform sample of 29 \Swift bursts rapidly observed by the Palomar 60-inch telescope through March 2008 (14 of which we classify as dark), all events have either a detected optical afterglow, a probable optical host-galaxy detection, or both.  Our results constrain the fraction of \Swift GRBs coming from very high redshift ($z>7$), such as the recent GRB 090423, to between 0.2--7 percent at 80\% confidence.  In contrast, a significant fraction of the sample requires large extinction columns (host-frame $A_V \gtrsim 1$ mag, with several events showing $A_V > 2-6$ mag), identifying dust extinction as the dominant cause of the dark GRB phenomenon.  We infer that a significant fraction of GRBs (and, by association, of high-mass star formation) occurs in highly obscured regions.  However, the host galaxies of dark GRBs seem to have normal optical colors, suggesting that the source of obscuring dust is local to the vicinity of the GRB progenitor or highly unevenly distributed within the host galaxy.
\end{abstract}


\section{Introduction}
% S1

The prevalence of ``dark'' gamma-ray bursts (GRBs) remains one of the most persistent mysteries of the field, twelve years after the discovery of GRB afterglows \citep{vP+1997,Costa+1997}.  While we now know that GRBs are frequently accompanied by extremely luminous afterglows (sometimes spectacularly so: e.g., \citealt{Akerlof+1999,Bloom+2009,Racusin+2008}) an optical detection is reported in only about half of cases since the launch of \Swift \footnote{See, for example, http://www.mpe.mpg.de/$\sim$jcg/grbgen.html and http://grbox.net}.

In contrast, an X-ray detection is nearly always reported for \Swift bursts \citep{Gehrels2008}.  Partly this is due to observational constraints:  the limitations of ground-based observing prevent a significant fraction of GRBs from being observed with terrestrial optical telescopes at all. 
Furthermore, the Ultra-Violet/Optical Telescope (UVOT; \citealt{Roming+2005}) 
on-board \Swift has a typical limiting magnitude 
that is shallower than the equivalent X-Ray Telescope (XRT; \citealt{Burrows+2005}) X-ray flux limit for a typical broadband afterglow spectrum, in particular when filters are applied.  Galactic extinction, stellar crowding, and proximity to the Sun or Moon, which do not significantly affect the X-ray band, also often complicate optical follow-up.  Estimates for the intrinsic frequency of optically dim GRBs vary and likely depend on the sensitivity of the detecting satellite, but for \Swift events \cite{Akerlof+2007} have estimated that approximately 30\% of GRBs have an optical magnitude $>22$ at only 1000 s after the trigger, and 15--20\% have an optical magnitude $>24$ at this time.  Detecting an optical afterglow from such an event requires a rapid response by a large-aperture telescope and is rare.

It is noteworthy that most of the conclusions about GRBs to date are based on a limited subsample of well-studied events that tends to exclude this large population of faint afterglows.  For example, evidence of a GRB-SN connection can only be established for known low-redshift events targeted for intensive photometric and spectroscopic follow-up (but c.f. \citealt{Levan+2005}).  Likewise, conclusions based on the nature of GRB host galaxies \citep{Bloom+2002,Fruchter+2006,Wainwright+2007} require accurate (generally sub-1\arcsec) positions.  Only a handful of pre-\Swift events without optical counterparts had sufficiently precise positions for later follow-up work of this nature.  Therefore the specific scrutiny of optically dark events is vital to understanding the entire GRB demography.

Key in the study of dark bursts has been the progression from a wholly observational definition of darkness to the physically motivated $\beta_{\rm OX}$ criterion of \cite{Jakobsson+2004}, who define a dark burst on the basis of the flux ratio between X-ray and optical bandpasses in the afterglow at 11 hr after the burst.  Here the parameter $\beta_{\rm OX}$ is the observed spectral index (defined using the convention $F_\nu \propto \nu^{-\beta}$) between the X-ray and optical bands, after correcting for Galactic extinction: $\beta_{\rm OX} = {\rm log}(F_{\rm X}/F_{\rm opt})/{\rm log}(\lambda_{\rm X}/\lambda_{\rm opt})$.  Jakobsson defines a dark burst as one with $\beta_{\rm OX} < 0.5$, motivated by the prediction from the synchrotron model in which, once the afterglow begins to fade, the intrinsic spectrum is given by $F \propto \nu^{-(p-1)/2}$ (for $\nu < \nu_c$) or $F \propto \nu^{-p/2}$ (for $\nu > \nu_c$), implying $\beta_{\rm OX} \geq 0.5$ if $p>2$.  \footnote{In addition to assuming $p>2$, this definition is meaningful only if the synchrotron model is assumed to be a complete description of the afterglow SED at these wavelengths.  We will make these assumptions throughout the paper.}

The availability and uniformity of X-ray follow-up in the \Swift era makes this definition of darkness particularly appropriate for a survey of \Swift bursts.  Even so, a purely optically defined criterion is still relevant: optical detection versus non-detection (rather than the flux ratio) is an essential factor determining the nature of further follow-up of the event: sensitive searches for host-galaxy dust, spectroscopic redshifts and measurements of the host ISM properties, and (to a lesser extent) accurate host identification require bright optical afterglows, making this likely the dominant selection bias affecting our current understanding of GRB afterglows and their origins.

The implications of dark bursts are potentially far-reaching, and the importance of folding them into our understanding of the GRB population as a whole is great, as---depending on the cause(s) of their optical faintness---there are reasons to suspect that their nature or environments may differ from those of the optically brighter GRBs which underpin our understanding of the field.  Some of the possibilities include \citepeg{Fynbo+2001}:

\begin{enumerate}
\item \textbf{Extinction.}  Dust in the GRB host galaxy (or elsewhere along the line of sight) can strongly obscure the rest-frame optical and ultraviolet light, dimming and reddening the afterglow \citepeg{Djorgovski+2001,Lazzati+2002,Reichart+2002}.  While previous (largely optically selected) samples have shown little evidence for widespread dust along GRB sightlines \citepeg{Kann+2006,Kann+2007,Schady+2007}, recent cases such as GRB 080607 ($A_V=3.2$ mag\footnote{Throughout this paper, $A_V$ refers to extinction in the host galaxy rest-frame $V$-band.}, \citealt{Prochaska+2009}) have demonstrated that very large dust columns can and do occur.  A bias against dusty galaxies in the current sample could easily mislead us in conclusions about, for example, mean GRB host metallicities and luminosities \citep{Fruchter+2006,Wolf+2007}.

\item \textbf{High redshift.}  GRBs have now been observed out to $z=8.3$ \citep{Tanvir+2009,Salvaterra+2009}.  At $z \gtrsim 6$, photons which would be redshifted into the optical bandpass are absorbed by neutral hydrogen in the host galaxy and IGM, suppressing the observed optical flux almost entirely \citep{GunnPeterson,Fan+2006}.  The redshift distribution of GRBs beyond $z \sim 6$ (and its implications on the star-formation history of the universe) cannot be observationally constrained without incorporating the dark burst population.

\item \textbf{Low luminosity.}  It is well-established \citepeg{Gehrels+2008,Nysewander+2009} that GRB fluence and afterglow flux are positively correlated (that is, underluminous bursts tend to also have underluminous afterglows).  Due to a wide distribution both in the depth of optical follow-up as well as in the gamma-ray fluence of observed GRBs, many nondetections could simply be attributed to follow-up that was not deep enough to constrain the predicted optical afterglow for a relatively faint GRB, without need to invoke absorption effects.

\item \textbf{Low-density medium.}  However, it is physically possible to have a energetic event without a luminous afterglow.  The afterglow phenomenon, which is thought to originate from shocks in the surrounding medium \citep{Paczynski+1993}, critically depends on the presence of circumstellar gas at sufficient density to excite bright synchrotron radiation.  GRBs exploding in galaxy halos or the intergalactic medium are predicted to have afterglows orders of magnitude fainter than those occurring in galactic disks \citepeg{kp2000}.
\end{enumerate} 

To some extent, these various possibilities can be disentangled via broadband observations of the afterglows of the events alone.  For example, a low-density medium will result in a dim afterglow at all wavelengths, extinction will suppress both the optical and the near-IR flux as well as soft X-rays (to different and characteristic extents), and a high redshift will suppress only the optical flux.   As a result, we will give attention in the subsequent discussion to the nature of the afterglows at all wavelengths.  However, extensive broadband follow-up is not always available (and the decision to trigger multi-wavelength observations carries its own selection biases), and in some cases the two possibilities are difficult to disentangle.

The remaining degeneracies can largely be broken via deep imaging of the host galaxy of a GRB.  In particular, high-redshift bursts should not have optically observable host galaxies, and the detection of a host can rule out the high-redshift hypothesis for that event.  Secondarily, study of the host galaxies themselves can determine whether our existing sample of pre-\Swift host galaxies is in fact typical, or if we are missing (for example) a large population of red, dusty ULIRGs.

\section{The Palomar 60-inch Sample}
% S2

The Palomar 60-inch telescope (P60; \citealt{cfm+2006}) is a robotic facility designed for moderately fast ($t \lesssim 3$\,minutes) and sustained ($R \lesssim 23$\,mag) observations of GRB afterglows and other transient events.  Fully operational since 2004 September, the P60 now routinely interrupts regular queue-scheduled observations in response to electronic notification of transient events.  The standard P60 response to \Swift GRB alerts results in a sequence of multi-color
($g R_{\mathrm{C}}i^{\prime}z^{\prime}$) observations for approximately the first hour after the trigger.  Subsequent observations are then triggered manually based on the properties of the afterglow in observations to that point.

The first catalog of P60 GRB observations was presented by \cite{Cenko+2009}.  The P60 follow-up program is fully robotic and the GRBs presented in that sample were selected entirely based on whether an event was rapidly followed-up.  P60 automatically follows up all \Swift GRB triggers that are observable, and therefore this catalog constitutes an effectively uniform sample of \Swift events to date, and should not be affected by any afterglow-related biases.  Other advantages of this population include a high afterglow detection efficiency 
(75\%, thanks to the relatively large aperture of the telescope and red filter sequence) and a large fraction with spectroscopic redshifts (60\%).

In total, the P60 sample contains 29 events (Table \ref{tab:sample}).  Of these, 7 were undetected with the P60 (to a typical limiting magnitude of $R > 20-23$, depending on conditions) at 1000 seconds.  No event that was undetected at 1000 s was detected at earlier times.  This is approximately consistent with the results of previous studies which have attempted to correct for the shallow follow-up of most \Swift GRBs in determining the true afterglow brightness function: in particular that of  \cite{Akerlof+2007}, which estimates (Figure 6 of that work) that 30\% of afterglows are fainter than 22nd magnitude at 1000 s.  These events are ``dark'' by the simple nondetection criterion, although the rapid response, large aperture, and nearly uniform depth of P60 makes a nondetection significantly more meaningful than is typical for \Swift bursts (many of which have no optical follow-up at all, or follow-up only from the UVOT and small-aperture ground-based telescopes.)  Four of the seven events have optical or infrared afterglows detected by other telescopes (typically with larger apertures and/or a redder wavelength response.)

We include a handful of additional events as ``dark'' via application of the $\beta_{\rm OX} < 0.5$ criterion of \cite{Jakobsson+2004}, though we apply it at 1000 s instead of 11 hr, given that late-time imaging is not always available and that our non-detection cutoff is also at 1000 s. \footnote{This involves some risks: there are occasional cases in which X-ray rebrightenings or strong spectral evolution is observed after 1000 s, indicating the contribution of additional prompt-like emission (X-ray flares) which have much harder spectra than a typical afterglow \citep{ButlerKocevski2007b} and could generate ``pseudo-dark'' events at early times which would look normal in later observations.  We will discuss the possibility of this contribution in the next section in the few cases where there appears to be evidence of extended activity at this time, but conclude that it is not a significant contaminant of our dark burst sample.}
There are 12 such events that satisfy this criterion: 5 of which are also P60 nondetections and 7 events which are detected by P60, but at a flux level that is less than a simple $\beta=0.5$ extrapolation of the 2 keV X-ray flux as determined by Table 3 in \cite{Cenko+2009}.
\footnote{Two events are listed with $\beta_{\rm OX} < 0.5$ at 1000s in \cite{Cenko+2009} which we do not include in our sample: GRB 050820A and GRB 071003.  In both cases, the \Swift XRT was not observing the source at 1000s and the actual spectral index at that time is unknown; the estimate in \cite{Cenko+2009} was based on an extrapolation from other epochs.  This is difficult, since GRB 050820A shows extensive early-time X-ray flaring while GRB 071003 experiences a dramatic rebrightening at around 1 day when XRT observations begin.  Late-time observations in both cases \citep{Cenko+2006,Perley+2009} show that the spectral index is quite normal at late times, strongly indicating that neither event is a genuine dark burst by either of our criteria (these are, in fact, among the two brightest bursts of the \Swift era.)}
Therefore our full ``dark'' sample defined by the union of both criteria consists of 14 events in all, approximately half of the P60 sample.  All 14 fields were imaged to deep limits at Keck Observatory, as discussed in the next section.

\begin{deluxetable*}{lccrlrl}
\tablecaption{P60 GRBs}
\tablewidth{0pt}
\tablehead{
\colhead{GRB} & \colhead{Fluence\tablenotemark{a}} & \colhead{X-Ray Flux\tablenotemark{b}} & \colhead{$R$\tablenotemark{c,d}} & \colhead{NIR\tablenotemark{c,d,e}} & \colhead{$\beta_{\rm OX}$\tablenotemark{f}} & \colhead{Reason for dark} \\
\colhead{} & \colhead{($10^{-7}$ erg cm$^{-2}$)} & \colhead{($\mu$Jy)} & 
\colhead{(mag)} & \colhead{(mag)} 
& \colhead{} & \colhead{classification} \\
}
\startdata
\multicolumn{7}{l}{Dark GRBs} \\
050412  &  6.18 &  0.27 &   $>21.4$  &            &  $<0.49$ & P60 nondetection     \\
050416A &  3.67 &  1.95 &     20.31  &            &  0.35    & low $\beta_{\rm OX}$ \\  
050607  &  5.92 &  0.45 &$\sim22.1$\tablenotemark{i}&            &$\sim0.33$& P60 nondetection     \\  
050713A & 51.1  & 14.51 &     18.45  &            &  0.31    & low $\beta_{\rm OX}$ \\
050915A &  8.5  &  0.72 &  $>20.7$   & $H\sim18$  &  $<0.44$ & P60 nondetection + low $\beta_{\rm OX}$ \\
060210  & 76.6  & 12.23 &      18.2  &            &  0.37    & low $\beta_{\rm OX}$ \\ 
060510B & 40.7  & 15.09 &$\sim20.4$\tablenotemark{j}&         &  0.04    & low $\beta_{\rm OX}$\\ 
060805A &  0.72 &  0.17 &   $>19.9$   &            &  $<0.76$ & P60 nondetection    \\
060923A &  8.69 &  0.92 &   $>22.0$\tablenotemark{k}& $K\sim18$\tablenotemark{k}  &  $<0.24$ & P60 nondetection + low $\beta_{\rm OX}$\\
061222A & 79.9  &  7.82 &   $>22.1$  & $K\sim18$  & $<-0.19$\tablenotemark{m} & P60 nondetection + low $\beta_{\rm OX}$\\
070521  & 80.1  &  4.40 &   $>22.9$\tablenotemark{l}& $K>18.7$   & $<-0.10$ & P60 nondetection + low $\beta_{\rm OX}$ \\
080319A & 48    &  1.19 &      20.43 &            &   0.41   & low $\beta_{\rm OX}$ \\
080319C & 36    & 11.68 &      18.32 &            &   0.36   & low $\beta_{\rm OX}$ \\
080320  &  2.7  &  1.37 &   $>21.0$  & $z'=20.0$  & $<0.31$  & low $\beta_{\rm OX}$ \\
\hline
\multicolumn{7}{l}{Other GRBs} \\
050820A & 34.4  &$\sim150$\tablenotemark{g}&   15.21  &            & $\sim$0.4&      \\
050908  &  0.51 &  0.12 &     19.17  &            & 0.91     &      \\
060110  & 15.7  &  7.42 &     15.46  &            & 0.80     &      \\
060502A & 23.1  &  1.22 &     19.50  &            & 0.53     &      \\
060906  & 22.1  &  0.20 &     18.84  &            & 0.88     &      \\
060908  & 28.0  &  0.92 &     17.59  &            & 0.82     &      \\
070208  & 47.7  &  0.88 &     19.74  &            & 0.54     &      \\
070419A &  5.58 &  0.17 &     19.02  &            & 0.87     &      \\
071003  & 83    & --\tablenotemark{h}&     17.06  &            &--\tablenotemark{h}    &      \\
071010A &  2.0  &  2.11 &     16.18  &            & 0.89     &      \\
071011  &  0.22 &  8.06 &     16.42  &            & 0.66     &      \\
071020  & 23    &  6.91 &     17.66  &            & 0.52     &      \\
071122  &  5.8  &  0.34 &     20.02  &            & 0.64     &      \\
080310  & 23    &  2.19 &     16.88  &            & 0.79     &      \\
080319B &810    & 265.8 &     13.69  &            & 0.52     &      \\  % 20.2^h
\enddata
\tablenotetext{a}{$15-150$ keV; taken from the BAT GRB table.}
\tablenotetext{b}{Absorbed flux at 2 keV; calculated using the \Swift XRT Repository \citep{Evans+2007}.}
\tablenotetext{c}{Calculated at 1000 s.}
\tablenotetext{d}{Vega mag; corrected for Galactic extinction.}
\tablenotetext{e}{Specified only in the case of $R$-band nondetections.}
\tablenotetext{f}{Between $R$-band and 2 keV.  From \cite{Cenko+2009}, modified include deeper non-P60 upper limits (where available) and revised XRT light curves.}
\tablenotetext{g}{The XRT was not observing this burst at 1000 s, and earlier observations were dominated by rapid flaring (see footnote in text).}
\tablenotetext{h}{The XRT did not slew to this burst until 22000 sec after the BAT trigger (see footnote in text).}
\tablenotetext{i}{\cite{GCN3531}}
\tablenotetext{j}{Based on extrapolation from later times: the burst was not detected in $R$-band at 1000s.}
\tablenotetext{k}{\cite{Tanvir+2008}}
\tablenotetext{l}{Interpolated between P60 measurements and \cite{GCN6436}.}
\tablenotetext{m}{\cite{GCN5978}}
\label{tab:sample}
\end{deluxetable*}

\section{Observations}
% S3

\subsection{The Keck Imaging Campaign}
% S3.1

Since 2005, we have been acquiring deep optical imaging of GRB fields using LRIS on Keck I \citep{Oke+1995} as part of our ongoing Berkeley Keck GRB Host Project.  Primary goals of the survey include elucidating the origins of dark GRBs, studying the hosts of X-ray flashes (XRFs; \citealt{Heise+2001}) and short gamma-ray bursts (SHBs), constraining late-time supernova emission from bursts, and identifying the hosts of GRBs showing strong DLAs \citepeg{Chen+2009} or intervening Mg II absorption \citep{Pollack+2009}.  As part of this project, we conducted imaging of all 14 P60 ``dark'' bursts above.  These observations were supplemented in a few cases by additional imaging taken by the Caltech GRB group, also with LRIS.

Our typical imaging mode was with the $R$ and $g$ filters simultaneously with the D560 dichroic, but other setups were also frequently used, with varying exposure times and total integrations depending on the field.  Observations were conducted between 2005 and 2009 across 15 different observing runs.  Some of these nights were photometric, and the photometry was calibrated using Landolt standard fields (SA 92, SA 101, PG 2213, and Markarian A: \citealt{Landolt1992}).  Non-photometric nights were calibrated using Sloan Digital Sky Survey (SDSS) photometry of stars in the GRB field \citep{DR6}, or when unavailable, using calibrations from the P60 matched to USNO standards \citep{Monet+2003}.  One field (GRB 060210) was calibrated to USNO directly ($B2$, $R2$, and $I$ magnitudes), using an average color of $B-g \sim 0.4 \pm 0.3$ \citep{Jordi+2006} to convert $B$ to $g$.   A list of all observations and exposure times is presented in Table \ref{tab:obs}.

\begin{deluxetable*}{lllllllll}
\tablecaption{Keck Imaging Observations}
\tablewidth{0pt}
\tablehead{
\colhead{GRB Field} & \colhead{Obs. Date} & \colhead{Filter} & \colhead{Int.\tablenotemark{a}} & \colhead{Seeing} & \colhead{Cal. Sys.} & \colhead{Cal. Unc.} & \colhead{5$\sigma$ Limit\tablenotemark{b}} & \colhead{$E_{B-V}$} \\
\colhead{} & \colhead{(UT)} & \colhead{} & \colhead{(s)} & \colhead{(\arcsec)} & \colhead{} & \colhead{(mag)} & \colhead{(mag)} & \colhead{(mag)} \\
}
\startdata
050412  &  2007-12-13 & $g$ & 690  & 1.4 & SDSS\tablenotemark{c}     & 0.03 & 25.8 & 0.02 \\
        &             & $R$ & 600  & 1.2 &                           & 0.17 & 24.5 & \\
050416A &  2005-06-05 & $g$ & 960  & 0.9 & SDSS                      & 0.03 & 26.2 & 0.03 \\
        &             & $R$ & 960  & 0.9 &                           & 0.02 & 25.4 & \\
050607  &  2007-10-09 & $g$ & 960  & 1.0 & Landolt\tablenotemark{d}  & 0.3\tablenotemark{g} & 24.4 & 0.156 \\
        &             & $R$ & 870  & 1.0 &                           & 0.3   & 23.7 & \\
050713A &  2008-08-02 & $g$ & 990  & 0.8 & P60/USNO\tablenotemark{e} & 0.25 & 25.7 & 0.414 \\
        &             & $R$ & 870  & 0.7 &                           & 0.27 & 24.7 & \\
050915A &  2005-12-04 & $V$ & 2280 & 0.7 & Landolt                   & 0.05 & 25.8 & 0.026 \\
        &             & $I$ & 1539 & 0.8 &                           & 0.02 & 24.9 & \\
        &  2005-10-31 & $g$ & 1680 & 1.0 & P60/USNO                  & 0.25 & 25.5 & \\
        &             & $R$ & 1500 & 1.0 &                           & 0.35 & 24.5 & \\
060210  &  2007-08-13 & $R$ & 540  & 0.7 & USNO\tablenotemark{f}     & 0.35 & 23.6 & 0.093 \\
        &  2009-02-19 & $g$ & 1680 & 0.8 &                           & 0.35 & 24.4 & \\
        &             & $I$ & 1530 & 1.0 &                           & 0.14 & 23.5 & \\
060510B &  2006-05-31 & $g$ & 3840 & 1.4 & Landolt                   & 0.02 & 25.8 & 0.039 \\
        &             & $R$ & 3660 & 1.4 &                           & 0.02 & 25.5 & \\
060805A &  2008-02-12 & $g$ & 1080 & 1.0 & SDSS                      & 0.04 & 26.3 & 0.024 \\
        &             & $R$ & 1260 & 1.0 &                           & 0.10 & 24.8 & \\
060923A &  2007-04-16 & $V$ & 1560 & 1.4 & SDSS                      & 0.04 & 25.2 & 0.060 \\
        &             & $I$ & 1590 & 1.2 &                           & 0.06 & 23.8 & \\
        &  2007-08-12 & $B$ & 1500 & 0.8 &                           & 0.07 & 26.4 & \\
        &             &$RG850$&1500& 0.6 &                           & 0.09 & 23.6 & \\
061222A &  2007-07-18 & $V$ & 710  & 0.8 & Landolt                   & 0.05 & 24.7 & 0.099 \\
        &             & $I$ & 600  & 0.7 &                           & 0.05 & 23.7 & \\
        &  2007-08-12 & $B$ & 1500 & 0.7 & P60/USNO                  & 0.12 & 25.9 & \\
        &             &$RG850$&1500& 0.6 &                           & 0.27 & 23.6 & \\
        &  2009-05-31 & $H$ &  900 & 0.5 & 2MASS                     & 0.06 & 21.6 & \\
        &             & $K$ & 1800 & 0.5 &                           & 0.09 & 21.7 & \\
070521  &  2007-05-21 & $V$ & 1500 & 0.7 & SDSS                      & 0.05 & 24.8 & 0.027 \\
        &             & $I$ & 1500 & 0.8 &                           & 0.03 & 24.3 & \\
        &  2009-06-25 & $V$ & 1440 & 0.7 & SDSS                      & 0.05 & 26.2 & \\
        &             &$RG850$&1260& 0.8 &                           & 0.15 & 24.6 & \\
080319A &  2009-02-19 & $g$ & 1070 & 0.6 & SDSS                      & 0.07 & 26.4 & 0.015 \\
        &             & $R$ &  960 & 0.7 &                           & 0.04 & 25.0 & \\
080319C &  2009-02-19 & $g$ & 1530 & 0.9 & SDSS                      & 0.05 & 25.6 & 0.026 \\
        &             & $R$ & 1380 & 0.7 &                           & 0.13 & 24.5 & \\
080320  &  2009-02-19 & $g$ &  990 & 1.0 & SDSS                      & 0.18 & 25.8 & 0.014 \\
        &             & $I$ &  810 & 1.3 &                           & 0.09 & 24.1 & \\
\enddata
\tablenotetext{a}{Total integration time.}
\tablenotetext{b}{As measured over a 1\arcsec\ aperture and averaged over the field; not corrected for extinction.  $BVRI$ magnitudes are in the Vega system. The $RG850$ filter is calibrated to the SDSS $z$-band.}
\tablenotetext{c}{Sloan Digital Sky Survey: \cite{DR6}}
\tablenotetext{d}{\cite{Landolt1992}}
\tablenotetext{e}{P60 calibration, based on USNO B1.0 catalog \citep{Monet+2003}}
\tablenotetext{f}{Direct calibration to USNO B1.0 catalog.}
\tablenotetext{g}{The two standard star observations during the 2007-10-09 run are not consistent with each other, indicating that this night may not have been photometric.}
\label{tab:obs}
\end{deluxetable*}

Images were reduced using standard techniques using a custom Pyraf script written in Python.  Astrometry was conducted relative to USNO-B1.0 astrometric standards.  In cases where we detected the optical afterglow with P60 (and, in one case, with the robotic infrared telescope PAIRITEL: \citealt{Bloom+2006}) these early-time images were registered and aligned with the Keck data to determine the most accurate possible afterglow position relative to the host-galaxy candidates.

\subsection{Host Identification}
% S3.2

Until recently, the same biases that made pre-\Swift host searches difficult without optical positions have applied to \Swift as well: early XRT positions were accurate to only 4--6\arcsec, an error region sufficiently large as to normally contain numerous faint galaxies.  However, by using optical sources to register the field \citep{Butler2007,Goad+2007,Evans+2009}, the \Swift XRT now routinely produces afterglow positions to better than 2\arcsec \ (90\% confidence).  
Furthermore, thanks to the proliferation of small- to medium-sized telescopes and the improving ability of larger apertures to respond relatively quickly, all but three of the $P60$-followed bursts in our sample are detected in the optical or IR.  In all cases where a host candidate is identified in or near the error circle, we follow the prescription in \cite{Bloom+2002} to estimate $P_{\rm chance}$.  Formally, this parameter is an estimate of the probability that one or more galaxies with an observed magnitude brighter than $m$ will be centered within a randomly chosen region on the sky with solid angle $\pi \theta^2$.  This probability is given by:

$$P_{\rm ch} = 1 - {\rm exp}(\pi \theta^2 \sigma_{\leq m})$$

Where $\sigma_{\leq m}$ is the average sky surface density of galaxies with apparent magnitude brighter than $m$, taken in this case from \cite{Hogg+1997}.  The values for $m$ and $\theta$ for each burst-host association are chosen as in \cite{Bloom+2002}, with two exceptions. Because we do not have access to space-based imaging and the size of a typical host galaxy is significantly smaller than the seeing disk, we conservatively use the visible extent of the optical disk in the ground-based imaging rather than the half-light radius.  We also use the 90\% confidence radius, rather than $3\sigma$, which is slightly less conservative.  We treat this value as an estimate of the probability that, for a given burst, the association with the nearest host galaxy is incorrect.

Some additional caution is warranted before interpreting $P_{\rm chance}$ this way.  In particular, this probability applies to a single event treated in isolation only: it is not necessarily appropriate for events chosen from a larger sample which includes both detections and nondetections (a shallow survey of a very large number of well-localized objects would find many individual low-$P_{\rm chance}$ galaxies even if the positions were chosen completely randomly).   Fortunately, in our case we identify good host galaxy candidates for most of our objects: 11 out of 14 fields contain at least one object with $P_{\rm chance} \leq 0.1$ consistent with the error circle.  Nevertheless, given the number of fields observed, we must recognize that the chance of a misidentification being present somewhere in the full sample is not insignificant.  A basic Monte Carlo analysis (including the nondetections) suggests that the probability of at least one chance coincidence being present in our host sample is an appreciable 48\%, and the probability of two or more is about 15\%.

The $P_{\rm chance}$ calculation also assumes that lines of sight toward GRBs, and in particular toward dark GRBs, are randomly sampled among all sightlines in the universe.  One possible interpretation of the overabundance of Mg II absorbers in GRB spectra relative to QSOs \citep{Prochter+2006} is that this assumption is incorrect and observed GRBs preferentially cluster along lines of sight near low-$z$ galaxies, perhaps due to gravitational lensing.  This interpretation is generally disfavored \citep{Prochter+2006}, and for the few cases of galaxy-associated Mg II systems in GRB spectra to date \citep{Masetti+2003,Jakobsson+2004b,Pollack+2009} there has been no clear demonstration that the number and offset distribution of these galaxies implies a significant excess of what is expected from chance.  Another possibility which could affect our results is if dark GRBs are due to extinction in unrelated field galaxies along the light of sight (rather than in the host galaxy) and therefore more likely to fall close to a line-of-sight galaxy: such an effect was studied as a possible interpretation of the GRB-QSO discrepancy in terms of a selection bias \citep{Sudilovsky+2007}.  Were this the case, dark GRB sight lines would be biased towards dusty foreground sources, and $P_{\rm chance}$ would be quite inappropriate for this sample.  However, given the highly confined distribution of dust in local galaxies and the observed density of galaxies on the sky, it would be surprising if a large fraction of GRB sightlines turned out to be attenuated; indeed, more detailed analysis by \cite{Sudilovsky+2009} has also recently shown that it cannot explain the GRB/QSO discrepancy either.  

For the purposes of this paper, we will assume no particular bias in GRB or dark GRB sightlines.  We shall return to this issue when discussing the implications of our large putative detection fraction in \S\ref{sec:conclusions}.

\subsection{Host Photometry}
% S3.3

\begin{figure*}
\centerline{
\includegraphics[scale=1,angle=0]{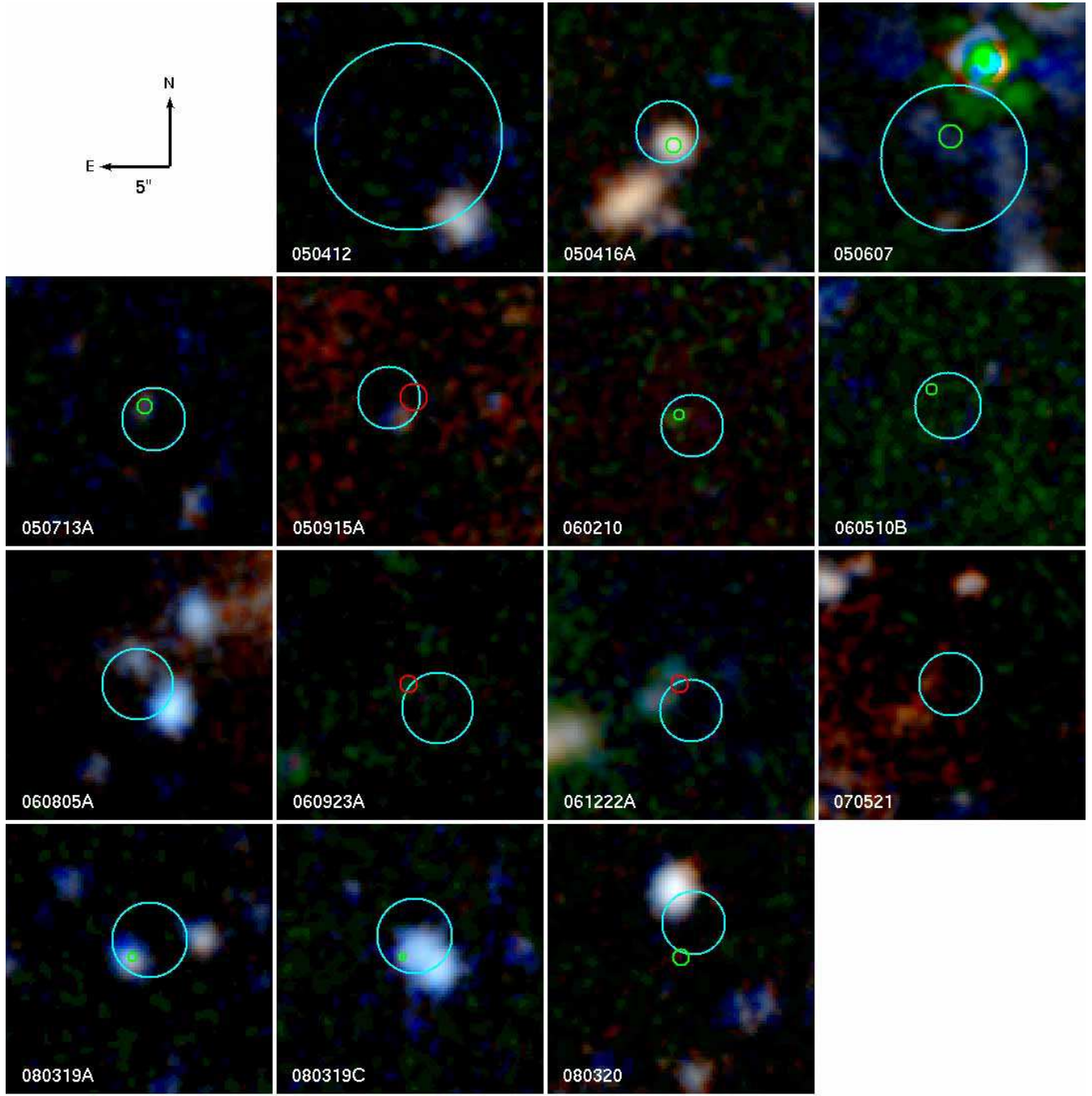}}
\caption{False-color mosaic of all 14 dark GRB host fields using our Keck LRIS imaging acquired between 2005 and 2009.  The 90\% confidence afterglow positions overplotted in each case.  X-Ray (XRT) error circles are cyan-colored, optical positions are green, and infrared positions are red.  All images are 11.8\arcsec\ on each side with north towards the top and east to the left. See Table \ref{tab:obs} for filter information.  In most cases images are constructed using two filters, with the green channel interpolated using a geometric mean.}
\label{fig:mosaic}
\end{figure*}

We used aperture photometry within IRAF to measure the flux of all candidate host galaxies, using a 1.0\arcsec \ aperture in all cases except for GRB 080319C, whose host is highly extended and a 2.0\arcsec \ aperture was used.   In a few cases, the afterglow position was within the outer point-spread function (PSF) of a bright star, which was subtracted prior to photometry using various techniques (depending on proximity and brightness, discussed below) to avoid the complication of a variable sky background as discussed in the relevant sections below.  The resulting aperture magnitudes are presented in Table \ref{tab:phot}.  A false-color mosaic of all imaging observations is presented in Figure \ref{fig:mosaic}.

% Figure 1

\subsection{Infrared Observations}
\label{sec:niri}
% S3.4

Two events in the sample, GRBs 061222 and 070521A, are of particular interest.  Both events were extremely X-ray bright, were not detected optically, and were observed at infrared wavelengths with large telescopes within a few hours after the burst.

GRB 061222A was observed \citep{GCN5978} to have a faint, fading IR afterglow.  We returned to this field on 2009-05-31 with NIRC on Keck I and integrated for 10 exposures of 100s each in $H$- and $K$-bands. (5 sec $\times$ 20 coadds).   Images were processed and stacked using a modified Python/pyraf script originally written by D. Kaplan and aligned to our LRIS imaging. The field was calibrated using a single Two-Micron All Sky Survey (2MASS) star within the NIRC field of view (2MASS J23530271+4632187).  We detect a faint source near the detection limit close to but not coincident with the infrared afterglow (likely a foreground galaxy very near the line of sight: see \ref{sec:061222A}).  No source coincident with the IR transient is detected.  Measurements and limits are reported in Table \ref{tab:phot}.

GRB 070521 was observed less than two hours after the burst by NIRI on Gemini-North \citep{GCN6450} and the lack of an IR detection imposes the deepest limit on a counterpart of any event in our sample.  The final UVOT-calibrated XRT position contains a red source (well-detected in $K$ and $H$, weakly detected in $RG850$, $I$ and $V$) near the eastern edge.  To rule out variability of this source, we acquired 24 $\times$ 60 s exposures in $K$-band and 18 $\times$ 60 s exposures in $H$-band on Gemini-North on 2009-02-01 (UT), 2.5 years after the burst.  The object is still detected in this imaging with no evidence for fading photometrically or in image subtraction of the frames.  The final IR photometry is presented alongside the optical photometry in Table \ref{tab:phot}.

\begin{deluxetable}{lllllll}
\tablecaption{Keck Host Galaxy Observations}
\tablewidth{0pt}
\tablehead{
\colhead{GRB Field} & \colhead{Obj.} & \colhead{$P_{\rm chance}$} & \colhead{Filter} & \colhead{Magnitude\tablenotemark{a}} & \colhead{AB magnitude\tablenotemark{b}} \\
}
\startdata
050412  & A  & 0.06 &  $g$ &$24.11 \pm 0.04$& $24.04\pm 0.04$ \\
        &    &      &  $R$ &$22.14 \pm 0.02$& $22.26\pm 0.02$ \\
        & B  & 0.45 &  $g$ &$25.82 \pm 0.18$& $25.75\pm 0.18$ \\
        &    &      &  $R$ &$25.08 \pm 0.33$& $25.20\pm 0.33$ \\
        & C  & 0.52 &  $g$ &$25.91 \pm 0.18$& $25.84\pm 0.18$ \\
        &    &      &  $R$ &$25.34 \pm 0.39$& $25.46\pm 0.39$ \\
        & D  & 0.40 &  $g$ &$> 27.05 $      & $> 26.98$       \\
        &    &      &  $R$ &$24.85 \pm 0.24$& $24.97\pm 0.24$ \\
050416A &    & 0.005&  $g$ &$24.11 \pm 0.03$& $24.00\pm 0.03$ \\
        &    &      &  $R$ &$23.10 \pm 0.02$& $23.19\pm 0.02$ \\
050607  &    &      &  $g$ &$> 25.0  $      & $> 24.44$       \\
        &    &      &  $R$ &$> 24.8  $      & $> 24.58$       \\
050713A &    & 0.006&  $g$ &$25.73 \pm 0.22$& $24.24\pm 0.22$ \\
        &    &      &  $R$ &$24.68 \pm 0.16$& $23.81\pm 0.16$ \\
050915A &    & 0.03 &  $g$ &$25.56 \pm 0.18$& $25.47\pm 0.18$ \\
        &    &      &  $V$ &$25.07 \pm 0.06$& $24.97\pm 0.06$ \\
        &    &      &  $R$ &$24.58 \pm 0.42$& $24.68\pm 0.42$ \\
        &    &      &  $I$ &$24.25 \pm 0.08$& $24.63\pm 0.08$ \\
060210  &    & 0.008&  $g$ &$> 25.6$        & $> 25.27 $      \\
        &    &      &  $R$ &$24.33 \pm 0.24$& $24.27\pm 0.24$ \\
        &    &      &  $I$ &$24.14 \pm 0.20$& $24.40\pm 0.20$ \\
060510B &    &      &  $g$ &$> 26.0        $& $> 25.86$       \\
        &    &      &  $R$ &$> 26.0        $& $> 26.07$       \\
060805A & A  & 0.05 &  $g$ &$25.46 \pm 0.04$& $25.37\pm 0.04$ \\
        &    &      &  $R$ &$24.45 \pm 0.07$& $24.56\pm 0.07$ \\
        & B  & 0.06 &  $g$ &$23.63 \pm 0.01$& $23.54\pm 0.01$ \\
        &    &      &  $R$ &$23.46 \pm 0.04$& $23.57\pm 0.04$ \\
        & C  & 0.22 &  $g$ &$24.63 \pm 0.04$& $24.54\pm 0.04$ \\
        &    &      &  $R$ &$23.97 \pm 0.05$& $24.08\pm 0.05$ \\
060923A &    & 0.06 &  $B$ &$> 27.2$        & $> 26.82$       \\
        &    &      &  $V$ &$26.19 \pm 0.30$& $25.98\pm 0.30$ \\
        &    &      &  $I$ &$24.67 \pm 0.24$& $24.99\pm 0.24$ \\
        &    &      &  $z$ &$> 25.23 $      & $> 25.12 $      \\
061222A & A  & 0.03 &  $B$ &$24.84 \pm 0.06$& $24.30\pm 0.06$ \\
        &    &      &  $V$ &$24.55 \pm 0.10$& $24.22\pm 0.10$ \\
        &    &      &  $I$ &$24.71 \pm 0.22$& $24.96\pm 0.22$ \\
        &    &      &  $z$ &$25.26 \pm 0.35$& $25.10\pm 0.35$ \\
        &    &      &  $H$ &$> 22.16 $      & $> 23.48$       \\
        &    &      &  $K$ &$> 22.23 $      & $> 24.03$       \\
        & B  & 0.02 &  $B$ &$24.41 \pm 0.04$& $23.87\pm 0.04$ \\
        &    &      &  $V$ &$24.30 \pm 0.07$& $23.97\pm 0.07$ \\
        &    &      &  $I$ &$24.21 \pm 0.13$& $24.46\pm 0.13$ \\
        &    &      &  $z$ &$24.92 \pm 0.26$& $24.76\pm 0.26$ \\
        &    &      &  $H$ &$21.84 \pm 0.30$& $23.16\pm 0.30$ \\
        &    &      &  $K$ &$21.91 \pm 0.29$& $23.71\pm 0.29$ \\
070521  &    & 0.10 &  $V$ &$26.29 \pm 0.20$& $26.18\pm 0.20$ \\
        &    &      &  $I$ &$25.08 \pm 0.33$& $25.46\pm 0.33$ \\
        &    &      &  $i$ &$25.25 \pm 0.17$& $25.20\pm 0.17$ \\
        &    &      &  $z$ &$24.10 \pm 0.16$& $24.04\pm 0.16$ \\
        &    &      &  $J$ &$22.52 \pm 0.20$& $23.40\pm 0.20$ \\
        &    &      &  $H$ &$21.58 \pm 0.09$& $22.94\pm 0.09$ \\
        &    &      &  $K$ &$20.95 \pm 0.10$& $22.78\pm 0.10$ \\
080319A &    & 0.03 &  $g$ &$24.63 \pm 0.03$& $24.58\pm 0.03$ \\
        &    &      &  $R$ &$23.85 \pm 0.06$& $23.98\pm 0.06$ \\
080319C &    & 0.01 &  $g$ &$23.08 \pm 0.03$& $22.99\pm 0.03$ \\
        &    &      &  $R$ &$22.22 \pm 0.03$& $22.32\pm 0.03$ \\
080320  &    &      &  $g$ &$>27.25$        & $> 27.20$       \\
        &    &      &  $I$ &$>25.3$         & $> 25.70$       \\
\enddata
\tablenotetext{a}{Not corrected for Galactic extinction.}
\tablenotetext{b}{Corrected for Galactic extinction.}
\label{tab:phot}
\end{deluxetable}

\subsection{Spectroscopy}
% S3.5

In several cases bright host candidates without afterglow absorption redshifts available were suitable for spectroscopic follow-up.  All spectroscopic integrations were conducted with longslit spectroscopy on LRIS, using the 400/8500 grating (red side) and 600/4000 grism (blue side) with the D560 dichroic, giving continuous spectroscopic coverage from the atmospheric cutoff to 9200\AA (using the old LRIS red chip) or out to 10400\AA \ (using the new LRIS red chip, which has greater quantum efficiency beyond 9000\AA and improved spectral range).  The exposures were reduced in IRAF using standard techniques and flux-calibrated using observations of standard stars BD+262606 and BD+174708 (red side) and BD+284211 (blue side) at similar airmass.  Absolute flux scales were then derived using the photometry derived from our previous imaging.  A summary of these observations is presented in Table \ref{tab:spec}.

\begin{deluxetable}{lllllll}
\tablecaption{Keck Spectroscopic Observations}
\tablewidth{0pt}
\tablehead{
\colhead{Field} & \colhead{Obs. Date} & \colhead{Exp.} & \colhead{Air-} & \colhead{Slit} & \colhead{PA} & \colhead{$\lambda$} \\
\colhead{} & \colhead{(UT)} & \colhead{} & \colhead{mass} & \colhead{($\arcsec$)} & \colhead{($\deg$)} & \colhead{(\AA)}  \\
}
\startdata
050412  &  2007-12-13 & 2$\times$900  & 1.16 & 1.0 & 142.85 & 3500--9150 \\
060805A &  2009-06-25 & 2$\times$900  & 1.21 & 0.7 & 30.40 & 3500--10400 \\
061222A &  2007-10-09 & 2$\times$1800 & 1.12 & 1.0 & 142.19 & 3500--9350 \\
080319A &  2009-06-25 & 2$\times$900  & 1.31 & 0.7 & 105.10 & 3500--10400 \\
\enddata
\label{tab:spec}
\end{deluxetable}

\subsection{Photometric redshift limits}
% S3.6

Even in the absence of spectroscopy, it is possible to place limiting a redshift on host galaxy candidates using the color observed our optical imaging.  Absorption of host-galaxy continuum light from hydrogen gas in the ISM at either the Lyman break (912 \AA) or Lyman-$\alpha$ (1216 \AA; \citealt{GunnPeterson}) will strongly suppress the observed flux once these features enter the $g$-band at about $z=3.4$ and $z=2.3$, respectively, greatly reddening the $g-R$ color and allowing us to translate an observed color into a limiting redshift.   We assume a strongly star-forming galaxy template (the Irr template from \emph{hyperz} [\citealt{Bolzonella+2000}], which due to its intrinsic blueness provides the most conservative choice) with no internal extinction, then apply a simple IGM attenuation correction from \cite{Madau+1995} to measure how its observed $g-R$ color evolves with redshift.  At sufficiently high redshifts, the Lyman-$\alpha$ forest and Lyman break sufficiently redden the galaxy light enough to be inconsistent with observations, generating a simple limiting photo-$z$.  If the redshift is known or well-constrained, a similar procedure can also be used to limit the internal extinction $A_V$.

\section{Dark Bursts and Host Galaxies}
% S4

\subsection{GRB 050412}
% S4.1

% Figure 2
\begin{figure}
\centerline{
\includegraphics[scale=1,angle=0]{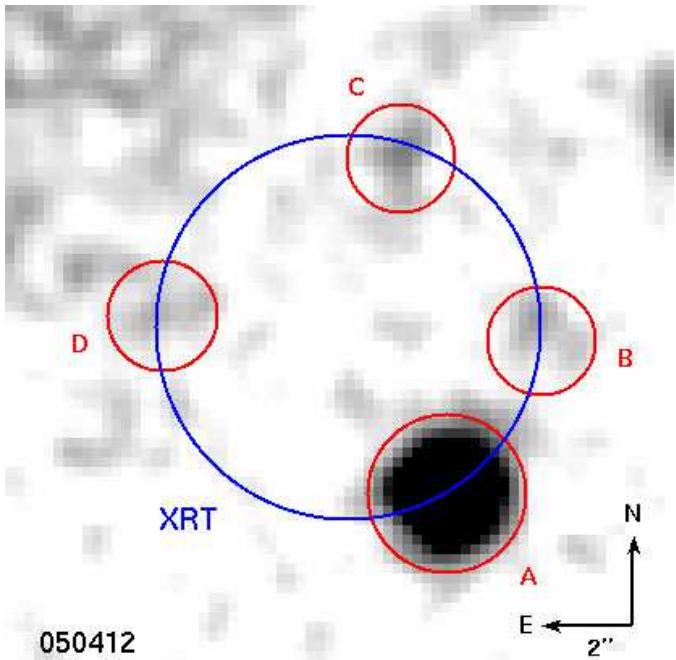}}
\caption{Keck/LRIS $R$-band image of the vicinity of GRB 050412 showing the four host galaxy candidates near the edge of the error circle.  The XRT error circle is relatively large; only object A is a statistically significant association ($P_{\rm chance} < 0.1$).
}
\label{fig:050412}
\end{figure}

The gamma-ray light curve of GRB 050412 shows no unusual features with a single peak and a long tail, and the prompt emission fluence (15--200 keV) is 9.6 $\times$ 10$^{-7}$ erg cm$^{-2}$ \citep{GCN3251},  near the median value for \Swift bursts.  The X-ray counterpart, however, is highly unusual.  
\Swift slewed to the position after only 99 s, and detected a fading source inside the BAT location.  However, the X-ray flux (after decaying slowly in the first 100 s of the exposure) plummeted abruptly starting around 300 s, with a decay index (defined by $F \propto t^{-\alpha}$) of $\alpha \sim 3$, and was not detected after about 1200 s \citep{Mineo+2007}.  A Chandra X-ray Observatory Target of Opportunity observation at 5 d \citep{GCN3291} failed to detect the counterpart.

This burst was relatively well-positioned for ground-based follow-up, and was tracked by several telescopes including P60, all of which failed to identify a fading counterpart.  Two additional observations deserve particular note: a Subaru integration at 2.3 hr which identified no afterglow to $R > 24.9$ mag \citep{GCN3263}, and rapid PAIRITEL follow-up which identified no infrared afterglow in observations starting at 175 seconds after the burst trigger.  Nondetections at such early times are rare among PAIRITEL-followed bursts (B. Cobb et al., in preparation).

\cite{Mineo+2007} speculate that the lack of afterglow flux of GRB 050412 might be the result of an extremely low-density environment suppressing the afterglow flux: a ``naked'' burst.  In this case, the X-ray afterglow is interpreted as being completely absent, with the sharply-decaying light curve attributed to photons from high latitude from the burst itself whose arrival at Earth is delayed by the curvature effect \citep{kp2000}.   A handful of other similar events exist in the literature, as discussed by \cite{Vetere+2008}.  However, such events are very rare (at most a few percent of \Swift bursts): plotting gamma-ray fluence versus X-ray flux \citep{Gehrels+2008,Nysewander+2009,Perley+2009}, 050412 is one of only a handful of outliers with extremely low X-ray to gamma-ray ratios.  The optical and IR nondetections are quite consistent with this picture---indeed, in terms of $\beta_{\rm OX}$, the available constraint of $\beta \lesssim 0.5-1.0$ is nothing unusual.  The darkness appears to be intrinsic, not due to absorption.

Presumably because of the weak and short-lived X-ray detection, the error circle of this event is  large.  A UVOT-corrected XRT position is not available, so the best available position is the one reported by \cite{Moretti+2006}: $\alpha = $ 12:04:25.19, $
\delta$ = $-$01:12:00.4 (unc. 4.2\arcsec)\footnote{All positional uncertainties in this paper are reported as 90\% confidence error circles.}.

A total of four sources are located within this region in our imaging, all of which are on the edge of the error circle (Figure~\ref{fig:050412}).  
The first object (A), which was reported by several groups in the GCN circulars \citep{GCN3243,GCN3244}, is bright ($R_{\rm AB} = 22.3 \pm 0.17$)\footnote{All reported host AB magnitudes and colors are corrected for Galactic extinction.  Afterglow magnitudes or those quoted from other sources are in the original reference system (Vega if $BVRI$, SDSS if $griz$) and are not corrected for extinction.} and very red ($g-R_{\rm AB} \approx 1.8$) with no clear emission lines over our spectral range in spite of its continuum brightness, which may suggest that it is an old galaxy with little star formation at moderate redshift (alternatively, it may also be an extremely luminous galaxy at $2.3>z>1.4$).  
Fitting line templates to the spectrum results in a best-fit redshift of $z\sim0.6$, but this is based on low-S/N absorption features.  In spite of the large XRT error circle, the brightness of the source gives a low $P_{\rm chance}$ of 0.06, making this a probable (though by no means definitive) host candidate.

Several additional, much fainter objects are also present near the edge of the XRT error circle.  One neighboring source (B) is not reported in any circular (likely because it was outside the original XRT error circle in the GCN circulars).  It is marginally detected in both filters ($R_{\rm AB} = 25.2 \pm 0.4$) and has a rather typical color.   A third source (C) was noted in the Subaru imaging of \cite{GCN3263} as being near the center of the original GCN XRT error circle.  It is weakly detected in our $g$-band imaging ($g = 25.84 \pm 0.18$) and marginally detected in our $R$-band imaging ($R_{\rm AB} = 25.46 \pm 0.39$) which is consistent with the report of a marginal detection with $R \approx 26.0$ in the Subaru imaging.   Finally, a fourth source (D) is at the top of the error circle and is detected with significance greater than $2\sigma$ in R-band only.  It is very red, with $g-R_{\rm AB} > 2$ mag.  All three of these additional sources have $P_{\rm chance}$ values of order unity.

The large XRT error circle, and the fact that all available host candidates are near its edge, makes host assignment particularly difficult in this case.  The only object whose presence in or near the error circle cannot be attributed to a chance alignment with probability of order unity is the brightest one (source A), but especially given that the original XRT position did not even include this source there is plenty of reason to be skeptical about the association.  If this is indeed the associated object, the combination of its red color, lack of lines, and perhaps even the fact that it is nearly outside the XRT error circle is particularly intriguing given the possibility of a very low circumburst density indicated by the X-ray light curve.

\subsection{GRB 050416A}
% S4.2

GRB 050416A (actually an XRF) is the second-lowest-redshift event in the P60 sample.  This GRB did have an optical afterglow that was detected by P60 and many other telescopes --- including the UVOT in its ultraviolet filters, suggesting that while this is a dark burst, it is perhaps a borderline case.  Indeed, in terms of $\beta_{\rm OX}$ (equal to 0.37 for this burst) this event is only slightly under the Jakobsson criterion.

The afterglow of GRB 050416A has been studied in detail by many authors \citep{Holland+2007,Mangano+2007,Soderberg+2007} and the presence of line-of-sight dust which may contribute to its optical faintness is, in principle, well-constrained.  \citet{Soderberg+2007} estimate $A_V \sim 0.87$ (using a Milky Way template), which compared to the majority of GRBs is already quite high, although \citet{Holland+2007} derive a significantly lower value of $A_V$ = 0.24.  This lower value is also favored by \citet{Kann+2007}.
%\footnote{\citealt{Holland+2007} estimate a much lower value of $A_V$ = 0.24, which may be due to inconsistent treatment of the broadband SED:  these authors fix the optical/IR spectral index to the X-ray value of $\beta_X \sim 1.0$ (unlike \citealt{Soderberg+2007}, they assume no cooling break) but do not fix its normalization, resulting in an SED that vastly underpredicts the observed X-ray flux.}

The host galaxy color is moderately red: $g-R_{\rm AB} = 0.8$; in part this is likely due to the presence of the 4000 \AA \ break between the $g$ and $R$ bands at the emission redshift of $z$ = 0.6535.  \cite{Soderberg+2007} detected the host in the \textit{HST} $F775W$ filter and estimate $I = 22.7 \pm 0.1$, corresponding to a significantly bluer color of $(R-I)_{\rm AB} \sim 0.15$.  Neither of these values constrain the host extinction strongly.  However, on the basis of the observed emission line ratio of $H_\gamma$/$H_\beta = 0.3\pm0.1$, they conclude that the host galaxy does likely harbor significant extinction.

\subsection{GRB 050607}
% S4.3

GRB 050607 is at the faint end of \Swift GRBs, with a fluence of $8.9\times10^{-7}$ erg cm$^{-2}$ \citep{GCN3525}.  Unfortunately, optical follow-up of this burst was greatly complicated by the presence of a bright ($R \approx 16$) star only 4\arcsec \ away from the burst location.  As a result, the P60 imaging of this burst is quite shallow, and no afterglow was detected in any filter.  However, even if stellar contamination were not a problem it is unlikely that P60 would have detected the afterglow, since much deeper observations with the KPNO 4m telescope \citep{GCN3531} do detect a transient with $I$ = 21.5 at 10 minutes, below the typical P60 limit even in an uncrowded field.  \cite{GCN3531} also note that the optical color is quite red, with $\beta_{\rm opt} > 1.5$: suggesting either substantial dust extinction or a high redshift ($z$ = 3--4).

The bright nearby star that complicated the P60 followup causes substantial difficulties for host follow-up also.  The star is saturated in our imaging, making PSF subtraction difficult, and the crowded field leaves no bright isolated template stars with which to accurately measure the PSF.  We fit and subtract the PSF of the nearby star (excluding the saturated core) using galfit \citep{Peng+2002}, and identify no obvious source at the position of the optical transient.  Therefore we are unable to strongly distinguish between the extinction and high-redshift possibilities, though the $B$-band afterglow detection imposes a limit of about $z < 4$.

\subsection{GRB 050713A}
% S4.4

GRB 050713A is another well-studied burst---mainly at X-ray and higher energies \citep{Morris+2007,Guetta+2007,Albert+2006}, as  unfortunately the optical coverage is much more limited.  It is bright, near the top end of the \Swift sample in both gamma-ray and X-ray flux.  The associated optical afterglow, however, is quite faint:  RAPTOR triggered on this burst and observed the event towards the end of the gamma-ray emission, but even at that point the event was only marginally detected with a peak magnitude of $R \approx 18.4$ \citep{GCN3604}.  Several prompt-emission flares at this time are seen in the X-ray and not the optical, but even after the X-ray flaring subsides the optical-to-X-ray index remains shallow at $\beta_{\rm OX} \sim 0.3$.  Unfortunately, this afterglow was detected in only $R$ and $I$ filters\footnote{Detections in $JHK$ filters have been reported by \citealt{GCN3583} but the photometry has not been made public.} and as a result the optical slope is only poorly constrained ($\beta_{\rm opt} = 1.4 \pm 1.0$) and on its own does not constrain the redshift of or extinction towards this GRB.

The position of this GRB is within the outer halo of a extremely bright star (1.1\arcmin \ from HD 204408, $V\sim6.6$ mag).  As a result, the region of the GRB is mildly compromised by a variable background, which we remove by applying a median filter over the region of the image around the GRB position.  After this step a source coincident with the optical position is clearly visible in $R$ and marginally detected in $g$.  The color of $g-R_{\rm AB}$ = 0.4 $\pm$ 0.3 does not constrain the nature of the galaxy given the unknown redshift.  It does limit the redshift to $z < 3.6$, ruling out any contribution of Lyman absorption to the observed afterglow faintness.

\subsection{GRB 050915A}
% S4.5

GRB 050915A is genuinely dark by all definitions.  It was followed up rapidly by several instruments, but only detected by one: the robotic infrared telescope PAIRITEL, which marginally detected a transient in $H$-band only ($H = 18.25 \pm 0.16$).   This is contemporaneous with $R$ and $I$-band nondetections with the P60 that require an afterglow spectral index of about $\beta_{\rm opt} > 1.45$, outside the range observed for typical unextinguished afterglows but only weakly constraining on the rest-frame extinction without additional constraints on the redshift and spectral index.  Furthermore, although this is not a particularly bright event in X-rays or gamma-rays, $\beta_{\rm OX}$ is clearly below the canonical dark value of 0.5.  There is no evidence of X-ray flaring or a flat energy reinjection phase after about 100s.

A faint galaxy, previously discovered by \citet{Ovaldsen+2007}, is well-detected consistent with the XRT position in all filters in which it was observed ($g$, $V$, $R$, and $I$).  It is somewhat offset (by about 1.1\arcsec) from the IR position, although because of the relatively low-significance detection of the infrared afterglow the 90\% confidence circle is large and its edge skirts that of the optical disk.  While $P_{\rm chance}$ is still low (0.06), we admit that this is one of the more tenuous associations in the sample.

While the optical detection of the host alone rules out a high-redshift origin, VLT spectroscopy of this galaxy (P. Jakobsson et al. in preparation) has revealed a surprisingly low redshift of $z\sim0.4$, indicating an extremely underluminous system ($M_{V(AB)} \approx -17.4$) and requiring a significant (though not, in this case, particularly large) dust column to explain the redness of optical afterglow.  Consistency of the combined X-ray and optical data requires $A_V \gtrsim 0.5$ mag independent of extinction law.

The blue colors of this galaxy indicate a young population free of widespread dust (global $A_V \lesssim 1.0$ mag from our template modeling).  This limit is not inconsistent with the relatively modest minimum extinction inferred from the afterglow.

\subsection{GRB 060210}
% S4.6

GRB 060210 provides significant insight into the dark burst phenomenon.  The optical afterglow of this burst was fairly bright, but only in the reddest bands ($R$ and $I$), peaking around 19.5 mag at a relatively late time of 600s following an extended episode of X-ray and optical flaring.  Afterglow spectroscopy 
by \cite{GCN4729} confirmed that this is a (moderately) high-redshift event at $z = 3.91$, explaining the steep fall-off towards the optical bands.    In addition, there is significant evidence for high-redshift dust, given that even optical filters redward of Lyman-$\alpha$ are significantly suppressed \citep{Curran+2007}.  
\cite{Cenko+2009} estimate $A_V = 1.21^{+0.16}_{-0.12}$ mag (in agreement with \citealt{Kann+2007}), which at the burst redshift corresponds to $\sim4$ mag of extinction in the observed $R$-band using an SMC template.

We imaged the field on two occasions; a relatively short $R$ integration followed by deeper $g$ and $I$ observations.   Unsurprisingly, nothing is detected in $g$-band, which falls below the wavelength of Lyman-$\alpha$ and is likely to be heavily obscured.  However, a bright source is detected at the OT position in $R$ and $I$.\footnote{This is not the object mentioned in \citealt{GCN4753}, which according to that note is 2-3\arcsec north of the XRT position.   No source is detected at that position in our imaging.}  The offset between this object and the OT is less than 0.5\arcsec \ ($P_{\rm chance} < 0.01$) and the association is further bolstered by the $g$-band nondetection.  This therefore likely represents among the highest-redshift GRB host galaxies detected to date, as well as among the most luminous ($M_R < -20.2$ for a starburst template).  In spite of the optical extinction, redward of Lyman-$\alpha$ the color of the object is quite blue, with $(R-I)_{\rm AB} = 0.1 \pm 0.3$ (the large uncertainty is dominated by the poor calibration of this field using USNO standards).  Given that the $R$ and $I$ bands correspond to wavelengths well into the ultraviolet at this redshift (1300--1700 \AA) where dust absorption is extremely efficient, this suggests that the average observed extinction cannot be high, though given the lack of knowledge about the extinction law it is difficult to constrain this formally.  For an assumed SMC-like extinction law, the host extinction is $A_V = 0.25 \pm 0.25$, which is certainly much less than the inferred extinction from the afterglow.  

\subsection{GRB 060510B}
% S4.7

The spectroscopic redshift of this event ($z = 4.941$, \citealt{GCN5104,Price+2007}) is the highest in the sample and among the highest for any burst to date.  At this redshift the Lyman-$\alpha$ transition is shifted well into the optical band, and consistent with this the flux in the P60 $R$ and $i$ bands is strongly suppressed.  Blueward of $R$-band the OT is not detected.  Unfortunately, this is one of the few bursts which displays clear flaring activity in the X-ray band as late as 1000 seconds after the GRB, making a consistent estimate of $\beta_{\rm OX}$ difficult, though as measured in $R$-band the burst is clearly dark for almost any assumption of the X-ray afterglow behavior.

Optically, coverage of this burst was sparse, and both $R$ and $i$ filters are affected by Lyman-$\alpha$ absorption, making it difficult to estimate the extinction.  However, the $z-J_{\rm AB}$ color of $0.0 \pm 0.4$ (based on the $J$-band point of \citealt{GCN5101}) requires $A_V < 0.5$ for $\beta_{opt} > 0$ and SMC-like extinction.  In addition, the late-time $\beta_{\rm IR-X}$ (using the $J$-band point) is actually $\sim 1.0$ and entirely normal, giving further evidence that the extinction is negligible.  Because of the known high redshift, our integration on this source was particularly long (approximately one hour), though the quality of the images is poor due to bad seeing (1.4\arcsec).  No object was detected at the P60 position or anywhere inside of XRT and \textit{XMM} X-ray error circles in either the $R$ or $g$ filters to 26th magnitude. 

The host galaxy of this burst was imaged by the Spitzer Space Telescope in a study conducted by \cite{Chary+2007}, and successfully detected with a flux level of $0.23 \pm 0.04$ $\mu$Jy.  Our $g$-band nondetection can be interpreted as support of this association (a detection of a galaxy blueward of the expected Lyman break in or near the optical position would indicate that the Spitzer source was actually an intervening source at lower redshift).  Given the high redshift, the $R$ nondetection is not surprising either; our limit of $R > 26$ corresponds to a luminosity of $M_R > -20.5$, which is still consistent with the luminosities of the majority of GRB hosts which have been observed to date \citep{Fruchter+2006} and with the sub-$L_{*}$ nature of the reported Spitzer host \citep{Chary+2007}.

\subsection{GRB 060805A}
% S4.8

% Figure 3
\begin{figure}
\centerline{
\includegraphics[scale=1,angle=0]{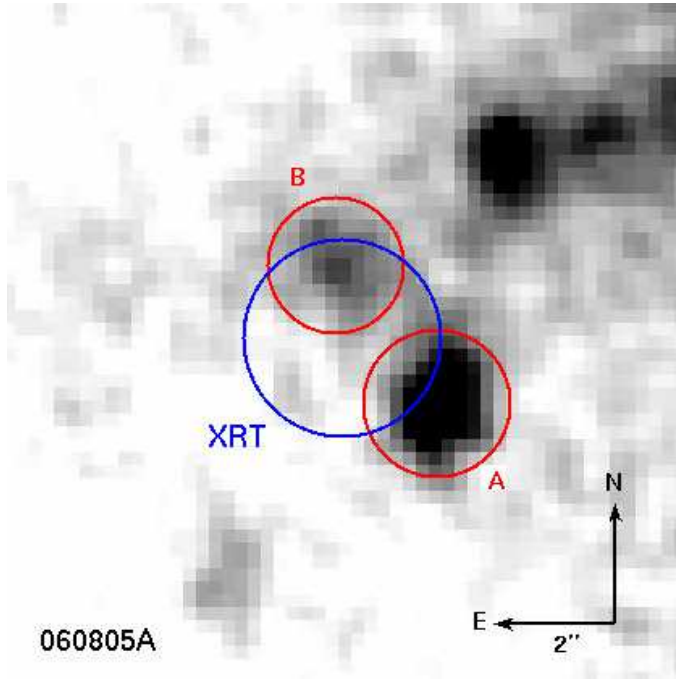}}
\caption{$R$-band imaging of the field around GRB 060805A.  Two galaxies are located inside the XRT error circle with additional objects nearby.}
\label{fig:060805A}
\end{figure}

GRB 060805A was an extremely faint \Swift burst, with a fluence value in the bottom 3 percent of all \Swift long GRBs (the burst was not detected at all above 100 keV).  The X-ray afterglow is extremely faint: $\approx 3 \times 10^{-4}$ mJy even at 100 seconds.

From this perspective it is no surprise that P60 (and all other optical instruments) failed to detect an optical afterglow, and indeed the limit on the optical to X-ray slope is effectively nonconstraining at $\beta_{\rm OX} <$ 0.7.   The low observed flux and fluence suggest an intrinsically low-luminosity event, though a typical-luminosity GRB at sufficiently high redshift could also appear faint simply because of its great distance.  Our imaging observations favor the former interpretation: two host galaxy candidates are present within the XRT error circle: one bright object (object ``A'' of Figure \ref{fig:060805A}, $R_{\rm AB} \sim 23.6$) at the southwestern edge and a second, fainter source (object ``B'', $R_{\rm AB} \sim 24.6$) slightly northeast of center.  The colors are significantly different:  the brighter source is blue with $g-R_{\rm AB} \approx 0$; the fainter one is redder with $g-R_{\rm AB} \approx 0.8$.  Unfortunately, we are not able to distinguish which is the correct host given the size of the XRT error circle.

Our spectroscopic observation of this source used a slit angle covering both sources (A and B).  Only the brighter object (A) shows a noticeable continuum trace in our 2D spectra.  No line features are observed over the spectral range down to the atmospheric cutoff; the nondetection of Lyman alpha or associated absorption features implies approximately $z<1.8$.  The nondetection of Lyman alpha at the position of object A may impose a similar redshift constraint on this object also, but this conclusion is less robust.  The redshift limit implied by the $g-R$ color is $z<3.8$.

\subsection{GRB 060923A}
% S4.9

% Figure 4
\begin{figure}
\centerline{
\includegraphics[scale=1,angle=0]{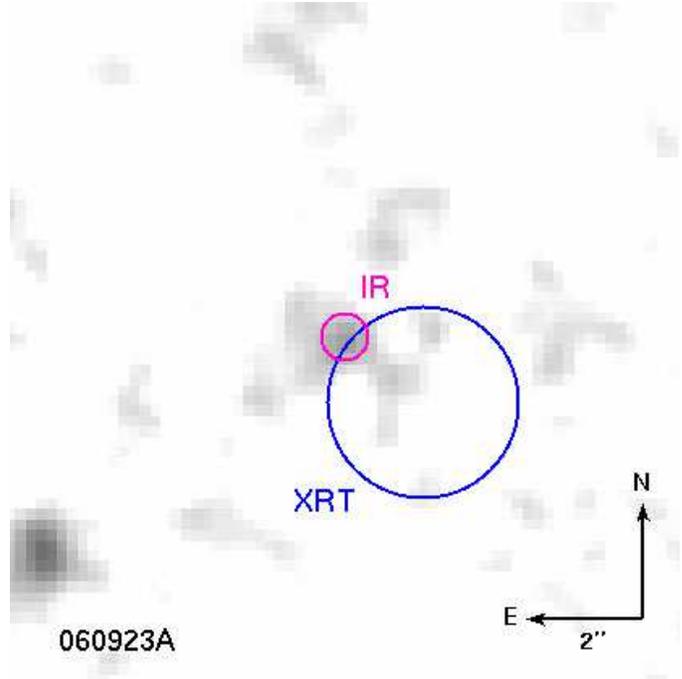}}
\caption{Stacked $V$- and $I$-band image of the field near GRB 060923A.  A faint galaxy is marginally detected coincident with the brightest, central region of the galaxy as also noted by \cite{Tanvir+2008}.   A projection from the galaxy appears to extend towards the southwest.}
\label{fig:060923A}
\end{figure}

One of the clearest examples of a dark burst in the sample is GRB\ 060923A.  Though not a particularly high-fluence event in gamma-rays or in X-rays, this burst was observed very early in the NIR ($<$ 1 hr) using UKIRT \citep{GCN5587} and shortly thereafter with both Keck and Gemini \citep{GCN5597}.  A transient was detected in $K$-band in all of these observations, but not in any bluer filter including $J$ or $H$.  One possible explanation for this would be an extremely luminous event at high redshift ($z > 15$).  However, later optical follow-up by \cite{Tanvir+2008} identified a host galaxy exactly coincident with the IR location, marginally detected in our imaging as well in $V$- and $I$-bands (Figure \ref{fig:060923A}).  It is not detected in $B$ or $RG850$.  \cite{Tanvir+2008} estimate that for $z = 2.8$ about $A_V \approx$ 2.6 would be sufficient to explain the inferred absorption.

The host galaxy is fairly but not remarkably red in the observed-frame optical: $(B-V)_{\rm AB} \gtrsim 0.5$ and $(V-I)_{\rm AB} = 1.0 \pm 0.4$.  A nondetection in $RG850$ rules out continuation of this trend further to the red, implying that the spectral energy distribution (SED) flattens towards the rest-frame optical, inconsistent with a highly dust-obscured source.  [\cite{Tanvir+2008} additionally report $(R-K)_{\rm AB} \sim 2.1$, which is not unusual for moderate-redshift galaxies.    We attempted to fit model SEDs using the combined $BVRIzK$ photometry, but due to the poor detections in all filters no reliable model converged.  Further, only a redshift limit of $z<4.4$ is possible from our photometry, though \citealt{Tanvir+2008} conclude that $z<4.0$ based on the combined properties of the X-ray and optical afterglows.]  Additional photometry will be necessary to reliably constrain the extent of extinction and other properties of the host, but as with most other galaxies in our sample the host-galaxy photometry does not demand large amounts of dust.

\subsection{GRB 061222A}
\label{sec:061222A}
% S4.10

%Figure 5
\begin{figure}
\centerline{
\includegraphics[scale=1,angle=0]{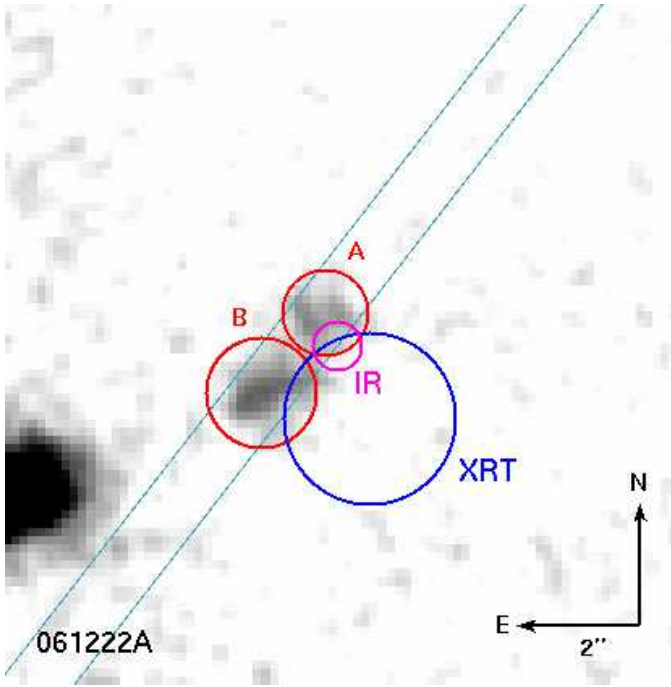}}
\caption{$V$-band image of the field near GRB 061222A.  Two objects with similar magnitudes and colors are slightly blended; only the northern object (A) is consistent with the Gemini infrared afterglow position \citep{GCN5975}. The position of the slit used in acquiring spectroscopy of the two sources is also shown.}
\label{fig:061222A}
\end{figure}

% Figure 6
\begin{figure*}
\centerline{
\includegraphics[scale=1,angle=0]{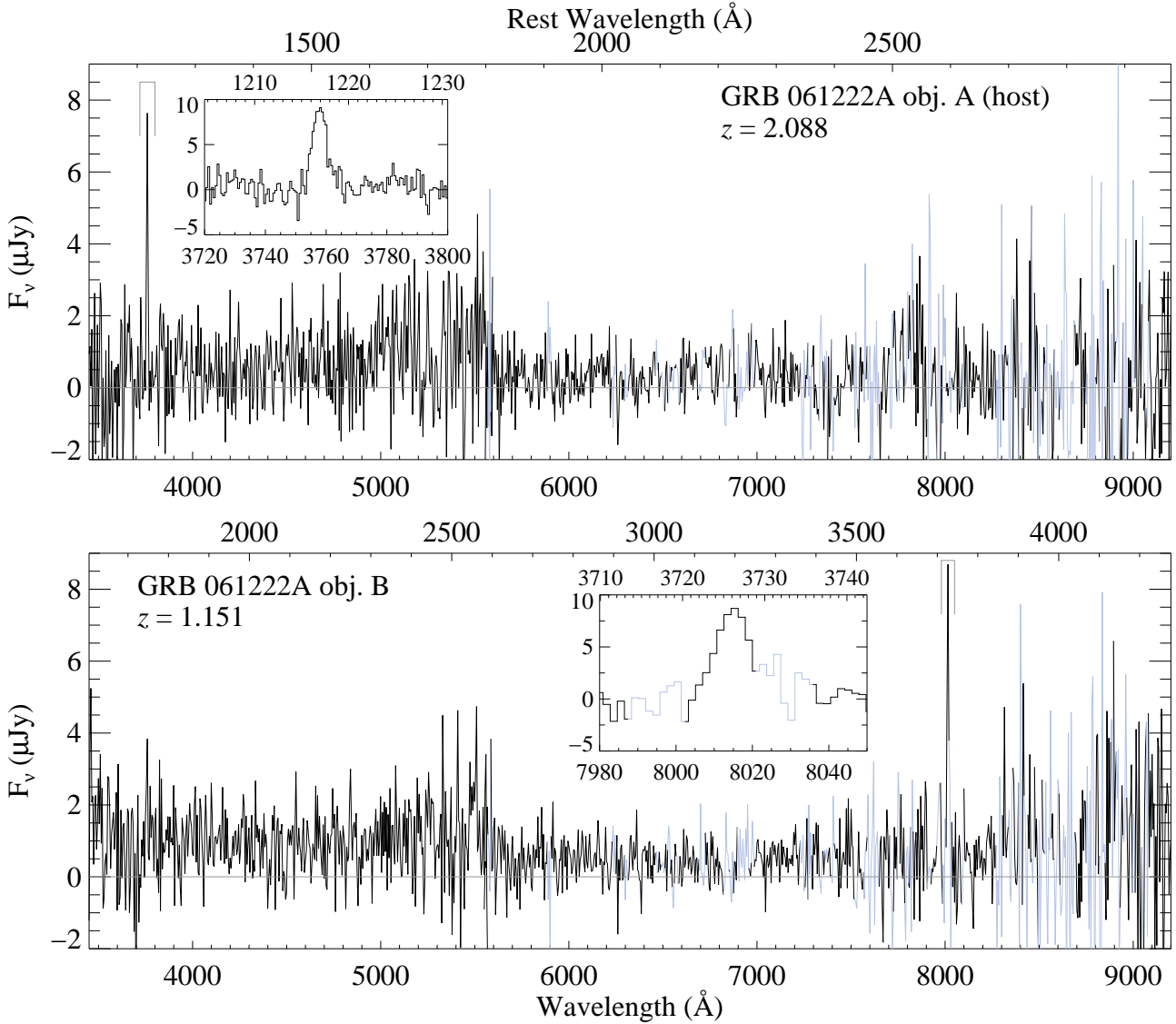}}
\caption{LRIS red-side spectrum of the host galaxy GRB 061222A and a nearby (in projection) object placed along the slit (Figure \ref{fig:061222A}) with insets showing detected emission lines.  A strong line is observed in the host galaxy (object A; top) at 3758\AA\ which we interpret as Lyman-$\alpha$ at a redshift of $z=2.088$.  No other objects are observed over our spectral range.  Despite the small offset and similar broadband color, object B is not at the same redshift.  No flux is observed at the location of the putative Lyman-$\alpha$ line; instead, we detect a single line at 8015\AA \ which we interpret as the [OII]3727 doublet at a redshift of $z=1.151$.}
\label{fig:061222Aspec}
\end{figure*}

At high energies, GRB 061222A is among the brightest events in the sample.  The gamma-ray light curve contains numerous separate pulses and extensive flaring out to $\sim$100 s, and the X-ray flux is also bright, well-detected by the XRT out to $10^6$ s.  As measured at $\sim 11$ hr the X-ray flux from this event is in the top 2\% of all \Swift GRBs.

Several other telescopes in addition to the P60 observed this event at early times, generally obtaining relatively shallow limits.  However, NIRI was triggered at Gemini in $K$-band only \citep{GCN5975}, and a faint source was identified that later faded, confirming this to be an infrared afterglow \citep{GCN5978}.  Unfortunately no deep imaging was acquired in other filters.  However, this event was also detected in radio using the VLA \citep{GCN5997}.

Two blended but seemingly distinct sources are observed near the afterglow position (Figure~\ref{fig:061222A}): one (source A) coincident with the IR transient and a second (source B) offset by about 1\arcsec \ to the southeast.  We identify the former as the host galaxy.   The two objects have similar colors, though photometry is complicated by the close blending, especially in the redder filters where neither object is well-detected.  Both galaxies are quite blue, with $(B-V)_{\rm AB} \sim 0.0$ mag, $(V-I)_{\rm AB} \sim 0.5$ mag, $(I-z)_{\rm AB} \sim 0.3$ mag.  Only object B is detected in the infrared, but both galaxies are clearly very blue in IR colors as well:  for object A, $(I-K)_{\rm AB} < 1.0$ mag; for object B, $(I-K)_{\rm AB} = 0.8 \pm 0.3$ mag.

Our LRIS longslit spectroscopic observation placed both objects along the 1\arcsec \ slit for two exposures of 1800s each.  The telescope was dithered 5{\arcsec} between the exposures.   The blue-side exposure was reduced normally, though the severe fringing on the red side was only removed effectively by subtracting the two exposures, which cleanly removed the fringe pattern.  We extracted spectra separately for both sources (A and B) along the slit near the afterglow position.  Interestingly, despite similar colors these galaxies are not at the same redshift.  The fainter, northern object (A), which we identify as the host galaxy, has a strong emission line at 3758 \AA.  No flux is observed at this position in the southern object (B).  At the same time, between two sky lines on the red side another bright emission line is clearly observed at 8014~\AA\, in this case consistent only with the position of object B.  The spectra and putative lines of both objects are shown in Figure \ref{fig:061222Aspec}.

The strong line in the blue part of the host-galaxy spectrum strongly suggests Lyman-$\alpha$ at a redshift of $z = 2.088$.  An alternate possibility is [OII] at $z=0.008$, but this would require an extraordinarily small and underluminous galaxy as well as imply the presence of H$\alpha$ at 6617 \AA, which is not observed.  Galaxy B cannot be at this redshift---its solitary line, if interpreted as [OII], indicates $z=1.151$.  (Alternatively, the line could be associated with H$\alpha$ at $z=0.22$, but this would predict the presence of [OII] at 4550 \AA\ which is not observed.)

At the observed redshift, any suppression of the bright optical afterglow predicted by the bright X-ray counterpart must be due to dust extinction.  The darkness of this burst is truly extreme: even in the observed $K$-band, approximately 4 mag of extinction are necessary if we assume the minimum synchrotron intrinsic spectral index of $\beta_{\rm OX} = 0.5$. At the observed host-galaxy redshift of $z=2.088$, this corresponds to approximately $A_V > 5.0$ mag (nearly independent of the choice of extinction law).\footnote{It is conceivable that the foreground object may also contribute to the extinction, but the blue colors of both this foreground object and the host (which would be reddened by a similar degree as the afterglow) make it unlikely to be a large contributing factor to the large absorption demanded by the afterglow.}

Given the enormous amount of extinction inferred from the faint infrared afterglow, one might expect that the relative amount of extinction in the observed optical bandpasses should be even greater---yet the host candidate is relatively bright ($V \sim 24$ mag) and extremely blue, showing no signs of reddening at all: the broadband color strictly limits the host-galaxy $A_V < 0.5$ mag.

\subsection{GRB 070521}
\label{sec:070521}
% S4.11

% Figure 7
\begin{figure*}
\centerline{
\includegraphics[scale=1,angle=0]{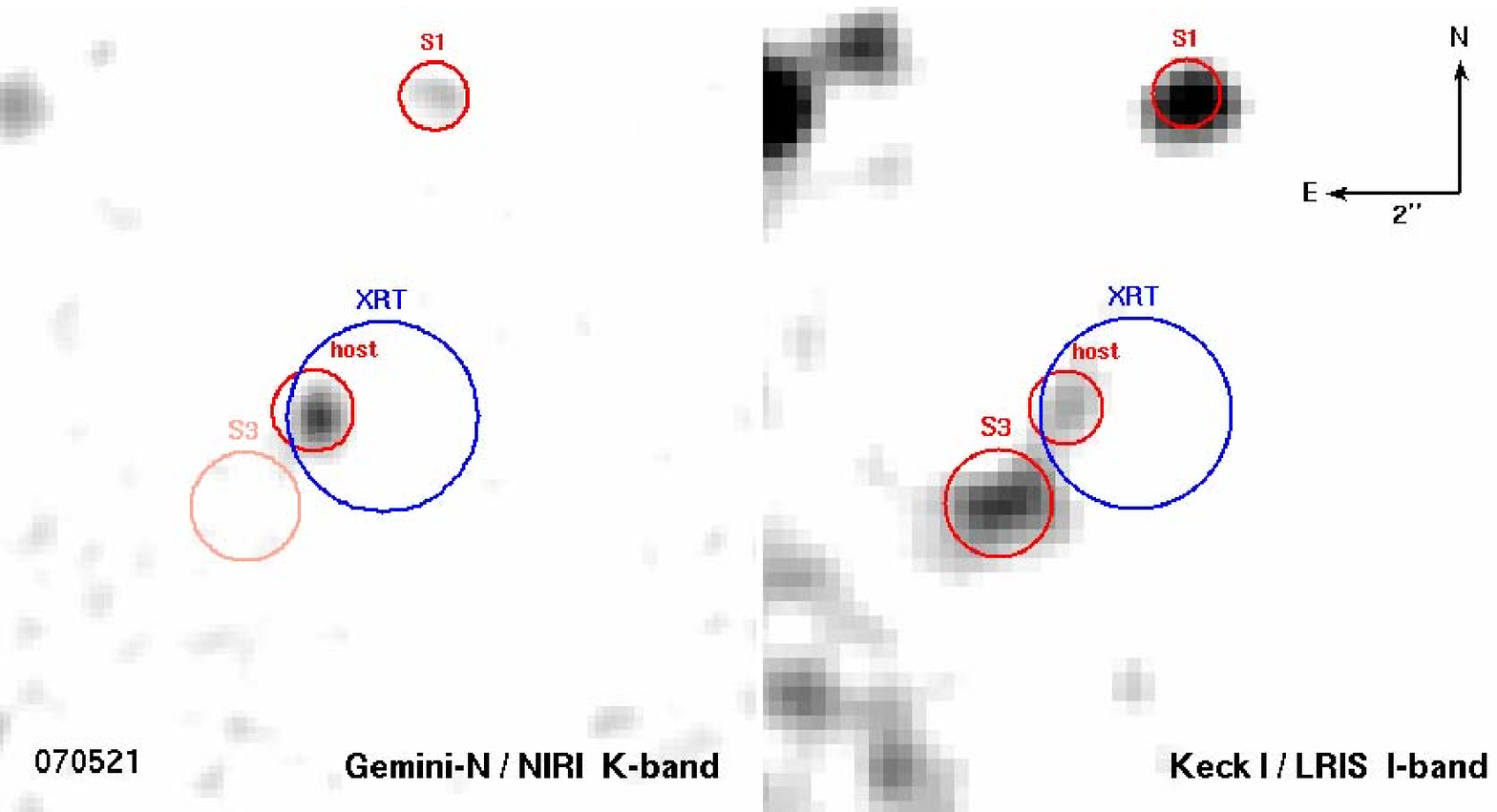}}
\caption{$K$-band imaging of the host galaxy of 070521 from NIRI on Gemini-North alongside $I$-band imaging from Keck.  S1 and S3 are nearby, unassociated objects that were proposed as possible hosts in the GCN circulars \citep{GCN6444,GCN6451}.}
\label{fig:070521}
\end{figure*}

Like GRB 061222A, GRB 070521 was a bright GRB with a bright X-ray counterpart.  In addition to the standard P60 follow-up, observations commenced at P200 within 1 hr \citep{GCN6436} and at both Keck and Gemini (including, in the latter case, $JHK_s$ IR imaging: \citealt{GCN6456}) within 2 hr.  As described in \S\ref{sec:niri}, no transient source within or near the XRT error circle was identified in any of this imaging despite the rapidity, depth, and relatively long wavelengths of these observations, making this burst the darkest in the sample.

In our observations, the most recent UVOT-enhanced XRT error circle includes a red, pointlike object near its eastern edge (Figure \ref{fig:070521}).  It is strongly detected in the NIR filters (except $J$, which was a relatively short exposure).  However, in $I$-band it is only marginally detected, slightly blended with another source located outside the XRT error circle, and was only detected in our $V$-band imaging after a second visit to the field: uniquely among the host-galaxy candidates in this sample, this object is quite red.  No other objects are present within the error circle at either optical or infrared wavelengths.

Thanks to the large suite of broadband photometry available for this object, we have been able to model the host SED and estimate an approximate photometric redshift.  Using the package \textit{hyperz} \citep{Bolzonella+2000}, the SED is well fit by a late-type galaxy template at a redshift of $z = 1.35^{-0.16}_{+0.32}$ with a stellar age of 360 Myr and an extinction of only $A_V$ = 0.4 mag.   The apparent redness is, therefore, more likely to be due to the presence of the 4000\AA \ break rather than dust: indeed, the $JHK$ SED redward of this break is quite normal.  Therefore, as with the other host galaxies in our sample, little dust extinction is demanded by the host data.

The amount of extinction required by the afterglow of this burst is as phenomenal as for 061222A.  Assuming an intrinsic afterglow $\beta_{\rm OX} > 0.5$ at $10^4$ sec, the deep Gemini limit requires an extinction of at least 4.7 mag in the observed $K$-band.  At the putative host redshift of $z=1.35$ this corresponds to a limit of $A_V > 9$ mag (over the 95\% confidence redshift range of $z=0.95-2.05$, the constraint is $A_V > 6$ mag).  A similarly large amount of extinction in the host SED is ruled out by our template modeling.

\subsection{GRB 080319A}
% S 4.12

GRB 080319A was a relatively bright GRB, though both its X-ray and optical afterglows are unremarkable, and the observational coverage sparse---likely as a result of the intense focus on GRB 080319B which occurred only 27 minutes afterward in the same part of the sky.  \Swift's initial slew to this burst was also delayed by 500 s due to an Earth constraint (in total, the XRT observed for only two epochs---at $\sim$ 1 ksec and briefly at 4 ks).  Optically, the afterglow is detected by P60 in $Riz$ filters and in a single epoch with the UVOT at approximately 600 s.  PAIRITEL also successfully detected the afterglow in $JHK$ before slewing to 080319B.  The IR color is also red and consistent with the optical color, for an overall optical-NIR spectral index of $\beta=1.5$.  This is suggestive of significant extinction.

A relatively bright galaxy is located coincident with the P60 optical afterglow position.  As with other galaxies in our sample, the optical color is not unusual ($g-R_{\rm AB} = 0.60 \pm 0.06$).  While this single color does not strongly constrain host extinction, as with other bursts the relative brightness of the host combined with the lack of obvious redness does not give any reason to suspect its presence.  Spectroscopy reveals no line features over our spectral range redward of the atmospheric cutoff, limiting the redshift to $z<2.2$.  At this redshift and assuming an intrinsic spectral slope $\beta < 1.2$, the lower limit on the extinction implied by the photometric SED is $A_V \sim 0.25$ mag (SMC extinction).  Any deviation from these assumptions (lower redshift, shallower intrinsic slope, or other standard extinction laws) would require additional extinction, implying a lower limit on the extinction of $A_V > 0.25$ mag.

\subsection{GRB 080319C}
% S 4.13

GRB 080319C was a bright, hard burst, and triggered several satellites in addition to \Swift including Suzaku, Konus, and Agile \citep{GCN7457,GCN7487,GCN7508}.  The afterglow is relatively unremarkable at late times, and was detected by the UVOT in filters as blue as $U$ and so clearly is not as ``dark'' as other objects in this sample ($\beta_{\rm OX}=0.36$).   The burst was in fact bright enough for an absorption redshift \citep{GCN7517} to be acquired, placing the event at $z$ = 1.95.  However, as is the case with the other bursts in the sample, the observed optical fluxes are suppressed relative to the X-ray flux 
and show evidence of reddening, which can be estimated with precision thanks to the known redshift and large numbers of filters ($A_V = 0.67 \pm 0.06$ mag, consistent with \citealt{Kann+2007}).   The optical and X-ray afterglows both show a dramatic flare around 200 s, after which the afterglow appears to decay relatively uniformly, though coverage is sparse.

The host galaxy of this event is remarkably bright: $R_{\rm AB} = 22.3$ mag.  In fact, this value is consistent with the reported P60 flux from observations taken the night after the GRB and was likely serendipitously detected even by this small-aperture telescope.  The relatively high redshift makes this particularly remarkable: the absolute magnitude of this galaxy for a flat-spectrum $k$-correction is $M_R = -22.6$ mag ($\sim$ 4 $L_{*}$ at $z \sim 2$; \citealt{Reddy+2008}), which would make it the second most luminous GRB host galaxy known (second to the even more remarkable host of GRB 081008, \citealt{GCN8372}, if its reported luminosity is real.).  The color is blue ($g-R_{\rm AB} = 0.33 \pm 0.05$ mag).  At the observed redshift this is not strongly constraining on the host dust: the 2175 \AA\ bump (if present---there is no evidence for it in the afterglow SED, though it is not strongly constrained) would shift into the $R$-band and as a result the broadband extinction is essentially gray around these wavelengths.

An alternate hypothesis for the galaxy detected in our imaging associates it with the $z=0.8104$ Mg II absorber in Table 54 of \cite{Fynbo+2009}, rather than the true host.  The apparent brightness of the object would be much less remarkable at this redshift, especially if the flux were combined with that of a true background host along the line of sight.   Spectroscopy or high-resolution imaging of this system will be needed to confirm or rule out this possibility.

\subsection{GRB 080320}
% S 4.14

GRB 080320 is a relatively faint \Swift burst with a mostly featureless light curve, though the X-ray light curve shows significant flaring ending at around 1000 s.  Due to the nearly full moon and the attention towards the previous night's GRB080319B, the optical afterglow was observed only sparsely.  This makes it difficult to accurately construct an SED of this event.  However, assuming no dramatic color changes or late-time optical flaring, all data are consistent with a very red afterglow color:  using contemporaneous or near-contemporaneous epochs we estimate $i-r > 1.1$ mag,  $z-i \approx 0.8 \pm 0.2$ mag, and $J_{\rm AB}-i \approx 2.2$ mag.   Alone, these observations are not sufficient to distinguish between a highly extinguished or high-redshift counterpart, though there is suggestion that both probably contribute: the SED is red across many filters, which is characteristic of extinction but less so of a Lyman break.  However, the $J$-band is probably not strongly suppressed relative to the X-rays ($\beta_{\rm OX} \approx 0.5$ as measured from $J$-band), and furthermore our early-time PAIRITEL limits on this event show no evidence for a bright $K$-band afterglow that may be expected if this redness carried into the optical.  The $i$-band detection imposes an upper limit on the redshift of $z<7$.

Consistent with this interpretation, we do not detect any host galaxy at the position of the optical transient to deep limits.  While in principle this could simply be the result of a low-luminosity host, the $N_H$ column measured by the XRT is relatively low in comparison with the dark bursts in our sample for which we infer large absorption columns, offering additional support of a moderately high-redshift origin (X-ray absorption is strongly wavelength-dependent, with the same column absorbing much more efficiently at lower energies: at higher redshift these lower energies are shifted out of the XRT sensitivity window and swamped by the Galactic absorption signature---see also \citealt{Grupe+2007}.)  Of course, a small host would predict a relatively low absorption column as well, though significant dust extinction in such a system would not be expected.  Nevertheless, we cannot rule out this scenario and can formally only place an upper limit on the redshift.

\section{Results}
% S5

\subsection{Redshift limits and the implications for high-$z$ GRBs}
% S5.1

An afterglow detection in any optical ($I$-band or blueward) filter immediately rules out a high-redshift origin.\footnote{Some measurable flux could be detected blueward of the Gunn-Peterson trough from a sufficiently luminous event at $5<z<7$, though such an event would show a clear photometric break.  We find no evidence for such an event in the P60 sample among GRBs with unknown redshifts.}  So does an optical detection of the host (assuming we have a proper identification).  Using the combination of these two factors we can place an upper limit on the number of bursts in our sample which could have originated from very high redshift ($z > 7$).

In fact, \emph{no} events of our sample are consistent with such a high-redshift origin.  If we assume our proposed host associations are all correct, \textit{all} 29 events in the P60 sample have either an optical transient or an optical host candidate, suggesting that---in spite of \Swift's sensitivity and customized trigger software---it detects few events beyond the range that has already been probed by optical spectroscopy.  Under this assumption, all events in the sample are constrained to $z<7$ and all but one (GRB 080320) to $z<5$.

\begin{figure}
\centerline{
\includegraphics[scale=0.8,angle=0]{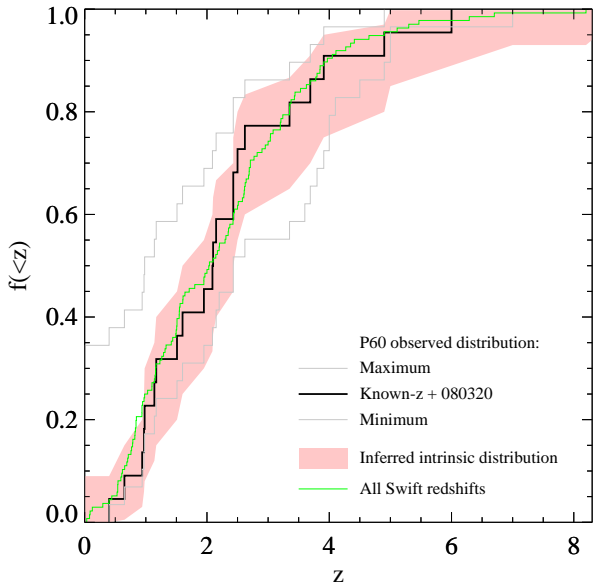}}
\caption{Cumulative redshift distribution of \Swift GRBs inferred from the uniform P60 sample.  The two solid gray lines show the absolute maximum and minimum observed redshift distribution in the entire sample (that is, assuming all GRBs with unknown redshift are at the maximum possible [see Table \ref{tab:zandext}] or the minimum possible [$z=0$] redshift).  
%% The dashed gray lines show the maximum and minimum for a more limited subsample, omitting GRBs without measured redshifts which do not have $\beta_{\rm OX} < 0.5$ (these non-dark are likely to have a similar redshift distribution as other events in the sample, though this is by no means assured).  The thick black line further omits three additional dark GRBs without measured redshifts but which are constrained to have $z<4$; GRB\ 080320 is conservatively assumed to be at $z=6$.  
The thick black line is the distribution omitting non-dark ($\beta_{\rm OX} > 0.5$) or potentially dark $z<4$ events with no measured redshift (these events are, as a population, not likely to significantly deviate from the redshift distribution of \Swift events in general) and conservatively assumes GRB080320 to be at $z=6$.  Based on this assumption, the salmon region then represents statistical limits on the cumulative fraction of \Swift GRBs originating at or below a given redshift as a function of $z$ permissible to be consistent with the observed distribution (10--90\% confidence limits).  The inferred distribution is generally consistent with the observed distribution of spectroscopic redshifts for all \Swift events to date, indicating that there are no strong redshift biases---except possibly at the highest-$z$ end, where the observed and intrinsic rate are not as well-constrained.}
\label{fig:zdist}
\end{figure}

Because the P60 sample is uniformly drawn from all \Swift events, we can convert this observational statement to a limit on the intrinsic high-redshift fraction among \Swift bursts.   We perform a simple Monte Carlo simulation in which 29 events are repeatedly drawn from a source population with the intrinsic high-$z$ fraction treated as an input parameter.  To convert this to a 90\% confidence upper limit, we then vary this input parameter until zero high-$z$ are events are drawn in exactly 10\% of the simulated 29-event samples (for $z > 7$) or zero or one event is drawn in exactly 10\% of the samples (for $z > 5$).  We conclude that, if all of our supposed associations are correct,  at most 13 percent of \Swift events are at $z > 5$ and at most 7 percent are at $z > 7$ to within 90\% confidence.  (This procedure can be generalized to lower redshifts also with appropriate assumptions---see Figure \ref{fig:zdist}.)

These estimates have neglected the possibility that some of our host associations may be chance alignments with foreground galaxies.  To take into account the possibility of foreground galaxy contamination, we assumed that 10\% of high-$z$ events in our simulation would be wrongly associated with a low redshift host ($P_{\rm chance} = 0.1$ is the largest observed in any of our possible host associations) and performed the simulation again, varying the true high-$z$ fraction until zero \textit{apparently} high-$z$ events are present in 10\% of the samples (for $z > 7$), or zero or one apparently high-$z$ events are present in 10\% of the samples (for $z > 5$).  In fact, this changes our constraints only slightly (by about one tenth of each percentage value).  We therefore conclude that, within 90\% confidence, \textbf{at most 14 percent of all \Swift GRBs are at $z > 5$} and \textbf{at most 7 percent are at $z > 7$}.

\begin{figure}
\centerline{
\includegraphics[scale=0.8,angle=0]{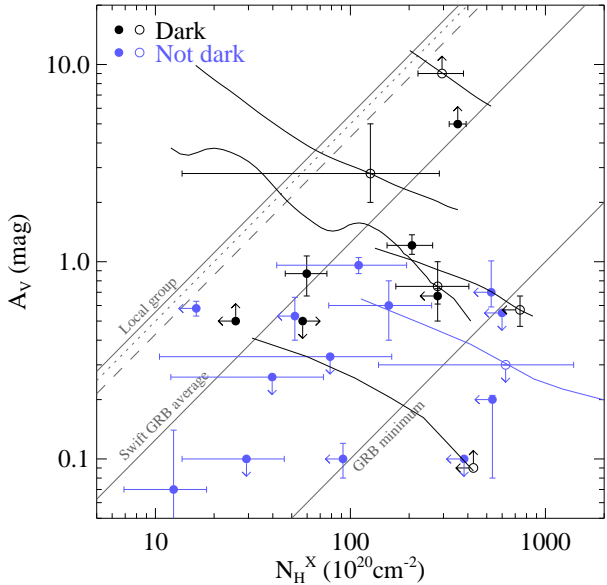}}
\caption{Rest-frame dust extinction $A_V$ versus $N_H$ as calculated from absorption in the X-ray spectrum due to light metal ions (assuming Solar metallicity) for 24 of 29 bursts in the P60 sample (five events, four dark and one non-dark, have been excluded due to the absence of meaningful constraints on either parameter).  Bursts with known redshift are shown as solid points; bursts with unknown $z$ are shown as open points at their most likely redshift (if only an upper limit is available, we plot the burst at a redshift of $z \sim 2$ or, in the case of GRB 080320, $z \sim 5$).   A ``track'' line then shows the evolution of the best-fit measurement or limit at different redshifts between $z=0.5$ and the maximum host or afterglow redshift in Table \ref{tab:zandext}.  The majority of events have a ratio of $A_V/N_H$ substantially lower than seen in Local Group galaxies, consistent with observations of other GRBs. (Milky Way, SMC, and LMC relations are plotted as lines using the values in \citealt{Schady+2007}, along with the average value for bursts in that paper and the minimum $A_V$/$N_H$ in the pre-\Swift sample of \citealt{Kann+2006}).  The high-$A_V$ events in our sample
($A_V > 2$), while not clearly inconsistent with the low $A_V/N_H$ relation observed previously, may suggest a trend towards more `normal' dust-to-gas ratios.}
\label{fig:avnh}
\end{figure}

As our most conservative assumption, we may choose to reject two host associations completely (in spite of the low $P_{\rm chance}$).  Specifically, suppose we reject two of the six host associations for events with no optical detection (for events with optical detections whether or not we have identified the host correctly does not significantly impact our conclusions about the redshift distribution)---that is, we assume a 33\% contamination rate, which is much higher than that anticipated by chance.  In this case, the data are consistent at 90\% confidence with up to 23\% of GRBs at $z > 5$ and up to 18\% at $z > 7$.  However, we point out that the ``weaker'' associations (where error circles and/or offsets are large: 050412, 050915A, 060805A) are consistent with simply being underluminous in all bands and no more likely to be at high-redshift than any other burst even if their claimed host galaxies are unassociated.  The one exception, GRB 070521, has independent confirming evidence for a  highly-absorbed, low-$z$ nature in the form of a  large X-ray $N_{\rm H}$ column (as do the statistically firmer associations of GRB 061222A and GRB 060923A; see also Figure \ref{fig:avnh}).

The recent detection of GRB 090423 at $z=8.2$ also allows us to place a (relatively non-constraining) lower limit on the high-redshift fraction.  While the P60 sample in this paper was cut off at the end of March 2008, P60 triggered rapidly on GRB 090423 and detected no afterglow to limits comparable to those discussed in this work.  GRB 090423 was the 42nd \Swift GRB on which P60 triggered rapidly.  We perform a simple Monte Carlo simulation in which bursts are sampled from an intrinsic population with a user-specified high-$z$ rate, which is varied until a high-$z$ event occurs as or earlier than the 42nd event 10\% of the time.  Only a rate of 0.2\% is necessary to fulfill this criterion.  Therefore, the detection of GRB 090423 requires (to $>$90\% confidence) only that a minimum of 0.2\% of all \Swift events originate from $z>7$, which is fully consistent with our maximum value inferred from the sample discussed in this paper.  Our overall constraint on the $z>7$ burst fraction for \Swift is therefore 0.2--7 percent (to within 80\% confidence).  This estimate is consistent with other recent observational limits on the high-$z$ fraction, such as that of \citealt{RuizVelasco+2007} ($\leq 19\%$ at $z>6$), \citealt{Grupe+2007} ($\leq7\%$ at $z>6$), and \citealt{Jakobsson+2005} ($7-40\%$ at $z>5$).

Our results strongly constrain some theoretical models of the evolution of the GRB rate with cosmic time.   For example, \cite{BrommLoeb2002} predicted that 50\% of all GRBs and 25\% of \Swift GRBs originate at $z>5$, which we rule out.  It is consistent with some more recent models that predict a low high-$z$ GRB rate based on star formation rate (SFR) models \citep{BrommLoeb2006,LeDermer2007}, luminosity indicators \citep{Li2008}, and limits on the GRB production efficiency of Population III stars \citep{Belczynski+2007,Naoz+2007}---though some of these models predict high-$z$ fractions close to our maximum value, which a larger sample may be able to confirm or refute.

\begin{deluxetable*}{lllllll}
\tablecaption{Redshift and extinction constraints on P60 GRBs}
\tablewidth{0pt}
\tablehead{
\colhead{} & \colhead{} & \colhead{Bluest} & \colhead{Bluest} & \colhead{} & \colhead{} & \colhead{$N_{\rm H}$ excess}\\
\colhead{GRB} & \colhead{$\beta_{\rm OX}$\tablenotemark{a}} & \colhead{AG det.} & \colhead{host det.} & \colhead{$z$} & \colhead{$A_{V}$\tablenotemark{b}\tablenotemark{c}} & \colhead{$(z=0)$\tablenotemark{d}} \\
\colhead{} & \colhead{} & \colhead{} & \colhead{} & \colhead{} & \colhead{(mag)} & \colhead{($10^{20}$ cm$^{-2}$)} \\
 }
\startdata
\multicolumn{7}{l}{Dark GRBs} \\
050412  & $<0.49$ & none &$g$?& $<$4.5?&       & $<$93.2      \\
050416A & 0.35    & UVW2 &$g$ & 0.6535\tablenotemark{e}&0.87& 24.0$\pm$6.0 \\
050607  &$\sim0.33$& $B$ &none& $<$4   &       & $<$15.0      \\
050713A & 0.31    & $R$  &$g$ & $<$3.6 &       & $<$28.3      \\
050915A & $<0.44$ & $H$  &$g$ &$\sim$0.4&$>0.5$& $<$14.4      \\
060805A & $<0.76$ & none &$g$ & $<$3.8 &       & $<$38.0      \\
060210  & 0.37    & $R$  &$R$ & 3.91   &1.21   & 8.7$\pm$2.1  \\
060923A & $<0.24$ & $K$  &$V$ & $<$4   &$\sim$2.5&22.1$\pm$9.2\\
061222A &$<-0.19$ & $K$  &$B$ & 2.088\tablenotemark{e}&$>5.0$&34.6$\pm$2.8 \\
060510B & 0.04    & $R$ &3.6$\mu$&4.941&$<0.5$ & $<$14.6      \\
070521  &$<-0.10$ & none &$V$ &$\sim1.35$&$>6$& 44.1$\pm$12.7\\
080319A &  0.41   & $r$  &$g$ & $<$2.2 &$>0.25$& $<$17.3      \\
080319C &  0.36   & $U$  &$g$ & 1.95   &0.67   & $<$32.6      \\
080320  &$<0.31$  & $i$  &none& $<$7   & -     & 8.7$\pm$ 3.3 \\
\hline
\multicolumn{7}{l}{Other GRBs} \\
050820A &         & UVW1 &$g$ & 2.615  &$<0.1$ & 3.4$\pm$1.5  \\
050908  &         & $V$  &    & 3.35   &$<0.55$& $<$19.3      \\
060110  &         & $R$  &    & $<$5   &$<0.3$ & ---          \\
060502A &         & $B$  &    & 1.51   &0.53   & $<$5.5       \\
060906  &         & $R$  &    & 3.685  &0.2    & $<$31.2      \\
060908  &         & UVW1 &$V$ & 1.884\tablenotemark{h}&$<0.1$ & $<$12.6      \\
070208  &         & $R$  &    & 1.165  &0.96   & $<$38.8      \\
070419A &         & $g$  &$r$?\tablenotemark{g}& 0.97   &0.70   & $<$35.8      \\
071003  &         & $U$  &    & 1.60435&$<0.26$& $<$13.9      \\
071010A &         & $g$  &    & 0.98   &0.60   & $<$37.0      \\
071011  &         & $V$  &    & $<$5   &       & $<$60.7      \\
071020  &         & $R$  &    & 2.145  &$<0.35$& $<$16.1      \\
071122  &         & white&    & 1.14   &0.58   & $<$10.6      \\
080310  &         & UVW1 &    & 2.43   &0.10   & $<$7.9       \\
080319B &         & UVW2 &$g$ & 0.937  &0.07   & 4.4$\pm$ 2.2 \\
\hline
090423\tablenotemark{f}&$<0.5$ & $J$  &none& 8.3    &0.1    & $<$10.6      \\
\enddata
\tablecomments{Properties of the full P60 sample (including non-dark bursts), modified from \cite{Cenko+2009} to include additional constraints based on the host galaxies and some additional afterglow data.  The redshift can be constrained in almost all cases.  See text for additional information.}
\tablenotetext{a}{Only listed for bursts identified as ``dark'' in the sample.}
\tablenotetext{b}{Extinction in the host-frame $V$-band along the line of sight inferred from the afterglow, generally assuming SMC extinction and $\beta_{\rm opt} \sim 0.6$.}
\tablenotetext{c}{$A_V$ references: \cite{Cenko+2009,Soderberg+2007,Covino+2008,Bloom+2009}}
\tablenotetext{d}{Equivalent $z=0$ hydrogen column density in excess of the Galactic value inferred from the X-ray spectrum, fit using the procedure of \cite{ButlerKocevski2007}. Only detections of $>2 \sigma$ excess are shown; other events are displayed as upper limits (see Figure \ref{fig:avnh} for a less conservative assessment of $N_H$ columns for bursts in the sample.)  All objects for which we infer large amounts of dust extinction in the optical band also have unambiguous detection of excess X-ray absorption columns; no event with low or negligible dust extinction shows this signature.  Since X-ray absorption is more efficient at low redshifts, 
this offers additional support to our association of these objects with moderate-$z$ hosts.}
\tablenotetext{e}{Emission-line redshift.}
\tablenotetext{f}{While not formally in our sample, the recent GRB 090423 is presented for reference as an example of a confirmed $z>7$ event.  Notably, this event has no host galaxy to $z>26$ and no significant excess $N_{\rm H}$ column \citep{Tanvir+2009}.}
\tablenotetext{g}{\citet{Dai+2008}.}
\tablenotetext{h}{Revised redshift from \citealt{Fynbo+2009}.}
\label{tab:zandext}
\end{deluxetable*}

\subsection{Constraints on dust extinction}
% S5.2

\begin{figure}
\centerline{
\includegraphics[scale=0.8,angle=0]{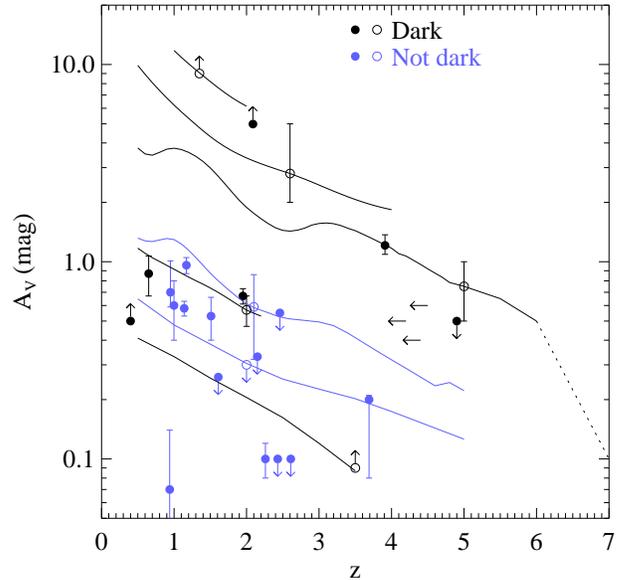}}
\caption{Constraints on rest-frame dust extinction ($A_V$) and redshift ($z$) inferred for all 29 bursts in the full P60 sample.  As in Figure \ref{fig:avnh}, bursts with known redshift are solid points; bursts with unknown $z$ are shown as a redshift ``track'' showing the evolution of the best-fit $A_V$ or limit with $z$; an open circle is plotted at a representative value.  For a few bursts $A_V$ is unconstrained; redshift limits are plotted as arrows at an arbitrary $A_V$ with no circle.  For clarity, the redshifts of two events have been adjusted slightly (less than 0.1) to prevent overlap of points.   All bursts are constrained to $z<7$ and all but one to $z<5$ (for the exception, GRB 080320, extinction is not constrained above $z \sim 6$, as shown by the dotted line).  However, many events show large extinction values, with a distribution skewed towards noticeably higher $A_V$ than previous, nonuniform samples \citepeg{Kann+2007}.}
\label{fig:avz}
\end{figure}

In Table \ref{tab:zandext} and in Figure \ref{fig:avz} we have summarized the extinction constraints 
derived based on our host-galaxy redshift constraints and the properties of the afterglow.  For most bursts in the full P60 sample, there is little extinction: the median $A_V$ is about 0.5, within the range of values typically seen in previous studies of optically well-studied bursts \citepeg{Kann+2006}.  However, large extinction columns are common: six bursts (out of 22 in which useful constraints can be derived) have $A_V > 0.8$ mag and three have $A_V \gtrsim 2.5$ mag.   
In comparison, only two events have $R$-band fluxes that are suppressed by hydrogen absorption at high redshift.  
Thanks to the uniform nature of this sample, we therefore are able to assert with reasonable confidence that the predominant cause of the dark burst phenomenon is dust extinction.  Even an extinction of $A_V \sim 1$ mag translates to large $R$-band extinction values at typical \Swift redshifts ($>3$ mag at redshifts of $z>2$).

Unfortunately, the nature of this dust remains a mystery.  The hosts of highly extinguished events tend to be unremarkable objects---often optically bright and with no evidence for large amounts of intrinsic reddening, and in a few cases with blue colors that appear to directly contradict the expectation of extremely red objects.  In no case are the optical colors indicative of a ULIRG-like highly extinguished object, which some theoretical models \citepeg{RR+2002} predict should be common among the GRB host population.  This result is not peculiar to our study: other dark burst hosts have, in the large majority of cases which have been studied to date, also been relatively blue objects without clear photometric evidence for extinction \citep{Jaunsen+2008,Rol+2007,CastroTirado+2007,Pellizza+2006,Gorosabel+2003,Djorgovski+2001}, though a few counterexamples of very red hosts do exist as well \citep{Levan+2006,Berger+2007}.

The results can be interpreted in several ways.  One possibility is that these apparently blue galaxies are concealing their true natures: if the distribution of dust is sufficiently patchy, it is conceivable that what looks like a normal object in the optical and NIR bands could harbor a massive starburst obscured from view by the same dust concealing the afterglow, allowing the (blue) emission from the optically thin regions to dominate the SED even if they contribute little to the total SFR.  Alternatively, there could be relatively little dust in the galaxy overall, but the GRB itself could be located deep within a relatively small dusty patch, such as a young molecular cloud, though this region would have to be sufficiently large to escape the destructive influence of the burst itself \citep{Waxman+2000}.  A third, more exotic possibility is that our templates for modeling high-redshift dust are incorrect, and high-redshift GRB hosts are dominated by grey extinction laws that redden their stellar populations relatively little \citep{Chen+2006,LiA+2008,Perley+2008}.  Unfortunately, the available data do not allow us to distinguish between these possibilities.

In any case, however, our results suggest that a significant fraction of GRBs (and, by association, of high-mass star formation) must occur within dusty regions not being probed by traditional optically-selected redshift surveys.  Based on our inferred distribution of $A_V$, we estimate that approximately $\sim$50\% of the rest-frame near-UV emission from \Swift GRB afterglows is absorbed by dust.  This value is quite similar to the fraction of obscured star formation inferred at high redshift based on far-infrared and millimeter studies \citep{CharyElbaz2001,LeFloch+2005} and may suggest that the potential for GRBs to serve as tracers of the high-redshift star-formation rate \citep{Blain+2000} is finally being realized.  Nevertheless, there is still need for caution: in addition to the possibility that the extinction may be a unique property of the GRB site hinted at by the blue observed colors of the host galaxies in our observations, there is evidence that metallicity or other biases result in a GRB host population strongly favoring subluminous galaxies in the local universe \citepeg{Modjaz+2008} and possibly at higher redshifts as well (\citealt{Fruchter+2006,LeFloch+2003}; cf. \citealt{Fynbo+2008})---which could skew the GRB positional distribution significantly away from that of high-$z$ star formation in general.

\section{Conclusions}
\label{sec:conclusions}
% S6

Twelve years after the discovery of a class of ``dark'' GRBs lacking optical afterglows \citep{Groot+1998}, we claim that the mystery surrounding the relative importance of the varying hypothesis for their apparent optical faintness is largely resolved.
Of 14 dark events (out of 29 events in the full P60 sample):
\begin{itemize}
\item {Seven or more events (070521, 061222A, 060923A, 060210, 080319C, 050416A, 080319A, plus likely 050915A and perhaps 050713A) are significantly suppressed by dust extinction (at least 1 magnitude in the observed $R$-band and typically much more) in their $z<4$ host galaxies.}
\item {Two events are probably suppressed due to Lyman absorption at redshift of $z>4.5$:  GRB 060510B at $z=4.941$ and GRB 080320 (at unknown redshift, but $z<7$).}
\item {Three events appear to be simply underluminous: not at high redshift, but because they were either intrinsically underenergetic (050607 and 060805A) or because little energy was coupled to the afterglow (050412, which may be a ``naked'' long GRB).  
Although no optical afterglow was detected for any of these events, they would not be classified as dark using the \cite{Jakobsson+2004} criterion.}
\end{itemize}

We conclude that the dark burst phenomenon is predominantly the result of extinction at moderate redshifts ($1<z<5$), with underluminous afterglows (otherwise normal events which are too faint for the sensitivity of a small telescope) also contributing significantly in an amount depending on the detection threshold---consistent with, but more constraining than, the results of pre-\Swift dark burst studies \citepeg{DePasquale+2003}.   In particular, a large fraction of high-redshift GRBs is not needed, and in fact is ruled out---providing observational evidence limiting the ability of Population III stars to efficiently produce GRBs and in agreement with most recent models of the high-redshift star formation rate.  Furthermore, our methods suggest that even if the discovery of very high-$z$ events continues to be extremely challenging (although the recent discovery of GRB 090423 at $z=8.2$ has now demonstrated that such events do exist and can be recognized), complementary host-galaxy searches can impose useful constraints on the high-redshift bursting rate free of selection biases, and we encourage continued host-galaxy follow-up of other medium-aperture robotic telescope samples and of dark bursts in general.  (A much larger, though not uniformly sampled, broadband survey of \textit{Swift}-era dark bursts is in progress.)

The location and nature of this high-redshift dust remains unknown: although our wavelength coverage is limited, no galaxy in our sample shows evidence of significant dust extinction.  In these cases, the line of sight to the afterglow must be passing through a much larger extinction column than the light from the observed young stars in the galaxy which dominate its rest-frame near-UV flux.  The solution likely requires that the dust is nonuniformly distributed---either closely linked with the GRB site itself, or widespread but sufficiently patchy to conceal its effects.  Although we cannot firmly resolve this question at this stage, it is clear that GRBs still have much to teach us about the structure of galaxies at high redshift and the importance of obscuration in the early universe.

Fortunately, the answers to these lingering questions may not be far off.  Longer-wavelength observations (near- and mid-infrared, sub-mm, radio) of these and other dark burst hosts would clarify the picture, piercing through the inferred dust screen or even detecting the reradiated emission from any postulated highly-extinguished population directly.   Such studies of the (limited) pre-\Swift dark burst host sample \citep{Barnard+2003,Berger+2003} indicate a population that differs little from GRB hosts in general---consistent with a patchy dust distribution in all GRB hosts, where the location of the GRB within its host (rather than the type of host) is the determining factor in the observed darkness of a given burst.  However, the obscuration rates derived from these radio and sub-mm studies are surprisingly high (typical radio/mm-derived host SFRs are 20--50 times the optically inferred values) and the non-detection of most such sub-mm sources in a recent survey by the \textit{Spitzer Space Telescope} \citep{LeFloch+2006} may call this conclusion of very high obscuration into some doubt.  The sample of \cite{LeFloch+2006} includes three ``dark'' GRBs, one of which (GRB 970828) is indeed associated with a strongly obscured galaxy (one of only two identified in their sample of 16 objects).  We suggest that more work in the long-wavelength regime is necessary to fully understand the nature of GRB host galaxies, especially of the most highly-extinguished events.

However, continued study in the optical band promises to be useful as well.  High-resolution space-based imaging could constrain the morphologies of dark GRB host galaxies, including any possible difference between the burst site and the rest of the galaxy.  (For example, studies of the host of pre-\Swift dark GRB 970828 seem to indicate a dust lane running through the afterglow position [\citealt{Djorgovski+2001}].)   The most luminous bursts are capable of shining through even very thick dust columns, allowing for detailed study of the material along their lines of sight.  Recently, spectroscopy and infrared photometry of the extremely bright GRB 080607 at $z=3.036$ revealed a strongly extinguished ($A_V=3.2\pm0.5$) event, showing a clear 2175\AA\ bump and an abundance of molecular and ionic lines associated with a nearby molecular cloud  with Solar-like metallicity \citep{Prochaska+2009}.  Similarly, \Swift bursts GRB 050401 \citep{Watson+2006} and GRB 070802 \citep{Kruhler+2008} were also ``dark'' events that were nevertheless sufficiently optically luminous to enable multiband photometry and optical spectroscopy, confirming the link between optical suppression (darkness), reddening, and dust absorption.  Such events, while rare, illustrate the need for continued observational effort on as many fronts as possible (including both spectroscopy and photometry, of both afterglow and host galaxies, and at all available wavelength regimes) to make further progress on the environments of gamma-ray bursts and their connection to star formation in the early universe.

\acknowledgments

J.S.B.'s group is partially supported by the Las Cumbres Observatory Global Telescope Network and NASA/\textit{Swift} Guest Investigator grant NNG05GF55G. SBC acknowledges generous support from Gary and Cynthia Bengier, the Richard and Rhoda Goldman fund, and National Science Foundation (NSF) grant AST–0607485.  Support for M.B. was provided by the W. M. Keck Foundation.  S.L. was supported by FONDECYT grant N$^{\rm o}1060823$

The W. M. Keck Observatory is operated as a scientific partnership among the California Institute of Technology, the University of California, and the National Aeronautics and Space Administration (NASA). The Observatory was made possible by the generous financial support of the W. M. Keck Foundation.  We extend special thanks to those of Hawaiian ancestry on whose sacred mountain we are privileged to be guests.

Gemini Observatory is operated by the Association of Universities for Research in Astronomy, Inc., under a cooperative agreement with the NSF on behalf of the Gemini partnership: the National Science Foundation (United States), the Science and Technology Facilities Council (United Kingdom), the National Research Council (Canada), CONICYT (Chile), the Australian Research Council (Australia), Ministério da Ciência e Tecnologia (Brazil) and Ministerio de Ciencia, Tecnología e Innovación Productiva (Argentina).

PAIRITEL is operated by the Smithsonian Astrophysical Observatory (SAO) and was made possible by a grant from the Harvard University Milton Fund, a camera loan from the University of Virginia, and continued support of the SAO and UC Berkeley. The PAIRITEL project are further supported by NASA/\textit{Swift} Guest Investigator grant NNG06GH50G and NNX08AN84G. 

This research has made use of the NASA/IPAC Extragalactic Database (NED) which is operated by the Jet Propulsion Laboratory, California Institute of Technology, under contract with the National Aeronautics and Space Administration.  This work additionally made use of data supplied by the UK Swift Science Data Centre at the University of Leicester.  We also acknowledge the hard work and efforts of the creators of other essential websites, in particular astrometry.net and GRBlog, which greatly assisted in this work.

We thank D. A. Kann and our anonymous referee for helpful comments and suggestions, and also thank Derek Fox as well as Pall Jakobsson and collaborators for sharing additional data on bursts within our sample.   Finally, we wish to acknowledge the hard work and dedication of the \Swift team, whose successful mission has made this study possible.

{\it Facilities:} \facility{Keck:I (LRIS)}, \facility{Gemini:South (NIRI)}, \facility{PAIRITEL ()},  \facility{PO:1.5m ()}

\end{document}